\documentclass[preprint2]{aastex}

\slugcomment{Re-submitted to ApJ in December 2011}

\shorttitle{Emission Mechanism of ``Green Fuzzies" }
\shortauthors{Takami et al.}

\begin{document}


%\bibliographystyle{astron}

\title{Emission Mechanism of ``Green Fuzzies''  in High-mass Star Forming Regions}


\author{Michihiro Takami\altaffilmark{1}, How-Huan Chen\altaffilmark{1,2}, Jennifer L. Karr\altaffilmark{1}, 
Hsu-Tai Lee\altaffilmark{1}, Shih-Ping Lai\altaffilmark{2},
Young-Chol Minh\altaffilmark{3}
}

\altaffiltext{1}{Institute of Astronomy and Astrophysics, Academia Sinica.
P.O. Box 23-141, Taipei 10617, Taiwan, R.O.C.; hiro@asiaa.sinica.edu.tw}
\altaffiltext{2}{Department of Physics, National Tsing Hua University, 101 Section 2 Kuang Fu Road, Hsinchu, Taiwan 30013, R. O. C.}
\altaffiltext{3}{Korea Astronomy and Space Science Institute, Daejeon 305-348, Republic of Korea}


\begin{abstract}

The Infrared Array Camera (IRAC) on the {\it Spitzer Space Telescope}
has revealed that a number of high-mass protostars are associated with
extended mid-infrared emission, particularly prominent at
4.5-$\micron$. These are called ``Green Fuzzy" emission or ``Extended Green Objects".
We present color analysis of this emission
toward { six} nearby ($d$=2--3 kpc) well-studied high-mass protostars and three candidate
high-mass protostars
identified with the Spitzer GLIMPSE survey. 
In our color-color diagrams most of the sources show a positive correlation
between the [3.6]-[4.5] and [3.5]-[5.8] colors along the extinction vector
in all or part of the region.
We compare the colors
with those of 
scattered continuum associated with the low-mass protostar L 1527,
modeled scattered continuum in cavities,
shocked emission associated with low-mass protostars, modeled H$_2$ emission
for thermal and fluorescent cases, 
and modeled PAH emission.
Of the emission mechanisms discussed above, scattered continuum provides the simplest explanation
for the observed linear correlation. In this case, the color variation within each object is attributed to
different foreground extinctions at different positions. 
Alternative possible emission mechanisms to explain this correlation may be  a combination of thermal and fluorescent
H$_2$ emission in shocks, {and a combination of scattered continuum and thermal H$_2$ emission,} but detailed models or spectroscopic follow-up are required
to further investigate this possibility. Our color-color diagrams also show possible contributions from PAHs
in {two} objects. However, none of our sample show clear evidence for PAH emission directly
associated with the high-mass protostars, several of which should be associated with ionizing
radiation. This suggests that those protostars are heavily embedded even at mid-infrared wavelengths.
\end{abstract}





\section{Introduction}

The UV radiation and outflows associated with high-mass protostars
hold keys for understanding high mass star formation and the formation
of clusters.  It has been debated for many years whether UV radiation
or stellar wind could stop mass accretion, preventing the formation of
high-mass stars \citep[see][for review]{Beuther07}. Of the scenarios
proposed to solve this issue, disk accretion is the most promising
\citep[see][for review]{Cesaroni07}. Furthermore, high-mass
protostars are often associated with clusters of low-mass
protostars. Energetic outflows and intense UV radiation from high-mass
protostars (or high-mass young stars) could affect their formation
and/or evolution, by destroying circumstellar material and/or
triggering new generations of star formation \citep[see][for review]{Stahler05}.

The Infrared Array Camera (IRAC) on the {\it Spitzer Space Telescope} has provided excellent capabilities for studying outflows and the effects of UV fields associated with star forming regions. Observations have shown the presence of extended infrared emission toward a variety of protostars and star forming regions. This emission is often
attributed to shocks associated with outflows, scattered continuum in the outflow cavities, or polycyclic aromatic hydrocarbons (PAH) excited by UV radiation. Such emission is expected to cover more than one or all filter bands (3.6, 4.5, 5.8, and 8.0 \micron) of IRAC \citep[e.g.][]{Reach06,SmithH06,Tobin07,Tobin08,Neufeld09,Takami10b}.
In Table \ref{line list} we summarize the bright H$_2$ lines and atomic/ionic lines which can be observed in IRAC bands.
These are often shown with conventional three-color images with different filters  \citep[in many cases blue, green, and red for 3.6, 4.5 and 8.0 \micron, respectively; e.g.][]{Noriega-Crespo04, Marston04, Rathborne05, SmithH06, Araya07, Kumar07, Shepherd07, Tobin07, Tobin08, Qiu08, Teixeira08, Cyganowski08, Cyganowski09, Cyganowski11, Chambers09, Morales09, Zhang09, Simpson09, Varricatt11}.

This three-color method provides useful diagnostics for the nature of these sources.
The PAH emission is brighter at 8.0-\micron~than the other bands \citep[e.g.,][]{Reach06}.
Hence it appears in red with the above color combination \citep[e.g.,][]{Shepherd07,Kumar07,Qiu08}. Other extended emission tends to appear ``green'', while stars are usually more ``blue'' as the flux is larger at shorter
wavelengths.  Studies of low-mass protostars show that emission from both shocks and scattered continuum appear green in three colors images \citep[see, e.g.,][]{Noriega-Crespo04, SmithH06, Tobin07,Tobin08, Teixeira08}. Although the adjustment of color contrast has been rather arbitrary in individual publications, such identification of emission
mechanisms seems to work fairly well (see also Section 2 for details).

A number of high-mass star forming regions are known to be associated with such ``green''  emission in the three color image with blue, green, and red for 3.6, 4.5, and 8.0 \micron~\citep[e.g.][]{Marston04,Rathborne05, SmithH06, Araya07, Shepherd07, Qiu08, Morales09, Simpson09, Varricatt11}. Candidates for high-mass star forming regions identified in this manner are called  ``Extended Green Objects" \citep[e.g.,][]{Cyganowski08} or ``Green Fuzzies" \citep[e.g.,][]{Chambers09,DeBuizer10}. H$_2$ and/or CO emission in outflow shocks are often regarded as the primary mechanisms to explain such emission \citep{Rathborne05, Araya07, Shepherd07, Qiu08, Morales09,Cyganowski08,Cyganowski09, Cyganowski11,Chambers09}. Observationally, it is in many cases based on comparisons with the studies of the well known massive protostellar outflow DR 21 \citep{SmithH06, Davis07} and the low-mass protostellar outflow in the HH 46/47 system \citep{Noriega-Crespo04}. \citet{SmithH06}, \citet{Davis07}, and \citet{Noriega-Crespo04} show that the 4.5-\micron~emission in these objects have a morphology very similar to H$_2$ emission at 2.12 \micron, indicating that the  ``green'' emission in these objects are associated with molecular shocks. The same trend has also been observed for several features observed in high-mass star forming regions \citep{Qiu08}.
Several Green Fuzzies are associated with molecular outflows observed at millimeter wavelengths \citep[e.g.,][]{Shepherd07,Qiu08,Cyganowski11}, Class I methanol masers, i.e., a tracer for molecular shocks \citep[][]{Cyganowski09,Cyganowski11}, or an ionized jet \citep{Araya07}. 

In contrast, in some cases other mechanisms such as scattered continuum can also be responsible for this emission.
\citet{Qiu08} attributed such emission in a few high-mass star forming regions to scattered continuum in outflow cavities, based on their morphologies and/or excess emission at 3.6-\micron.
\citet{DeBuizer10} have made ground-based infrared spectroscopy of two Green Fuzzy objects, and conclude that the knotty structures in one of the objects are dominated by thermal H$_2$ emission, while the other is associated with scattered continuum.
\citet{Varricatt11} shows that the bright part of the extended emission in the 4.5-\micron~and
%$K$-band
{2.2-\micron}
continuum positionally match fairly well in IRAS 17527-2439, suggesting the same origin. He shows that this object is also associated with a bipolar jet in H$_2$ 2.122 \micron~emission, but with different distribution from the other wavelengths.
\citet{SmithH06}  analyzed their results with spectra obtained by {\it the Infrared Space Observatory}, and showed that fluorescent H$_2$ emission also significantly contribute to this region in addition to shocked emission in the DR 21 outflow.

Clear understanding of the nature of the emission in individual objects would allow us to investigate the activity of the associated high-mass protostars in detail.
While infrared spectroscopy is a powerful tool for this purpose \citep{DeBuizer10}, it covers significantly limited areas of emission regions in many cases.
% Quantitative analysis of colors obtained by imaging observations would overcome this problem, allowing us to discuss the emission mechanism in the entire regions of our interest. On the other hand, if the emission is due to scattered continuum, spectroscopic follow-ups would allow us to access to a variety of emission absorption features, thereby allowing us to investigate mass accretion and ejection at the very close proximity (of an order of 10$^3$ K) of individual high-mass protostars \citep[e.g.,][]{Scoville83, Elias06} or their evolutionary sequence by observing photospheric lines \citep{Hosokawa09,Hosokawa10}
%
We therefore tackle this issue with the observed fluxes, flux ratios and color-color diagrams based on IRAC data, in
particular at 3.6, 4.5 and 5.8 \micron. We compare the colors of
Green Fuzzy emission with various cases, including
scattered continuum emission associated with a low-mass protostar  \citep[L 1527 ---][]{Tobin08}, modeled scattered continuum emission,
shocked emission in low-mass protostellar jets, thermal H$_2$ emission calculated by \citet{Takami10b},
and modeled PAH \citep{DL07} and fluorescent H$_2$ emission \citep{DB96}.
In Section 2 we describe our targets and data. In Section 3 we describe our analysis and results with flux ratio maps and color-color diagrams.
In Section 4, we perform comparisons between the observed colors and those for the various emission mechanisms described above. In Section 5 we discuss possible contributions from individual mechanisms. In Section 6 we briefly comment on the emission mechanisms not included in this study. In Section 7 we provide conclusions.

\section{Targets and Data}

Our targets are summarized in Table \ref{target list}. These include {six} nearby
($d$=2-3 kpc) high-mass protostars and three bright candidates for high-mass protostellar
objects reported by
\citet[][]{Cyganowski08}. The
selection of the high mass targets is based on Beuther \& Shepherd
(2005), who reviewed previous interferometric observations of
high-mass protostellar outflows at high angular resolutions. These
objects are categorized into high-mass protostellar objects without
any evidence for ionizing radiation (HMPO), hypercompact HII regions
(HC HIIs) and ultra-compact HII regions (UC HIIs).
%We select a few
%objects from each category: two HPMOs (IRAS 05358+3543, IRAS
%16547--4247); three HC H IIs (G 35.2--0.7 N, G192.16--3.82, IRAS
%20126+4104); and two UC HIIs (G 5.89--0.39 and W 75 N).
We select two HPMOs (IRAS 05358+3543, IRAS
16547--4247); three HC H IIs (G 35.2--0.7 N, G192.16--3.82, IRAS
20126+4104); and an UC HII (W 75 N).
The additional
three objects are selected from
the catalog of Cyganowski et al. (2008). In these objects the emission is
fairly extended so that we can apply analysis with color-color diagrams
at different positions (Sections 3 and 4).
These  targets are cataloged as ``likely massive young stellar objects
with outflows" by \citet{Cyganowski08}, based on the presence of extended 4.5 \micron~
emission without confusion from nearby point sources and/or
problematic saturation.
In addition to
the high-mass protostars, we also analyze the data for the L 1527 protostellar outflow, a
well-known low-mass protostar associated with scattered continuum
in conical bipolar cavities \citep[][]{Tobin08}.
{Some of the above regions show diffuse PAH emission with an external
origin. In each of these objects, the spatial variation of the PAH emission in
the field is much smaller than the flux from the object at 3.6, 4.5 and 5.8 \micron,
for which we apply the color-color analysis in Sections 3 and 4.}

\begin{figure*}
\epsscale{2.2}
%\epsscale{1.0}
\vspace{-1.5cm}
\plotone{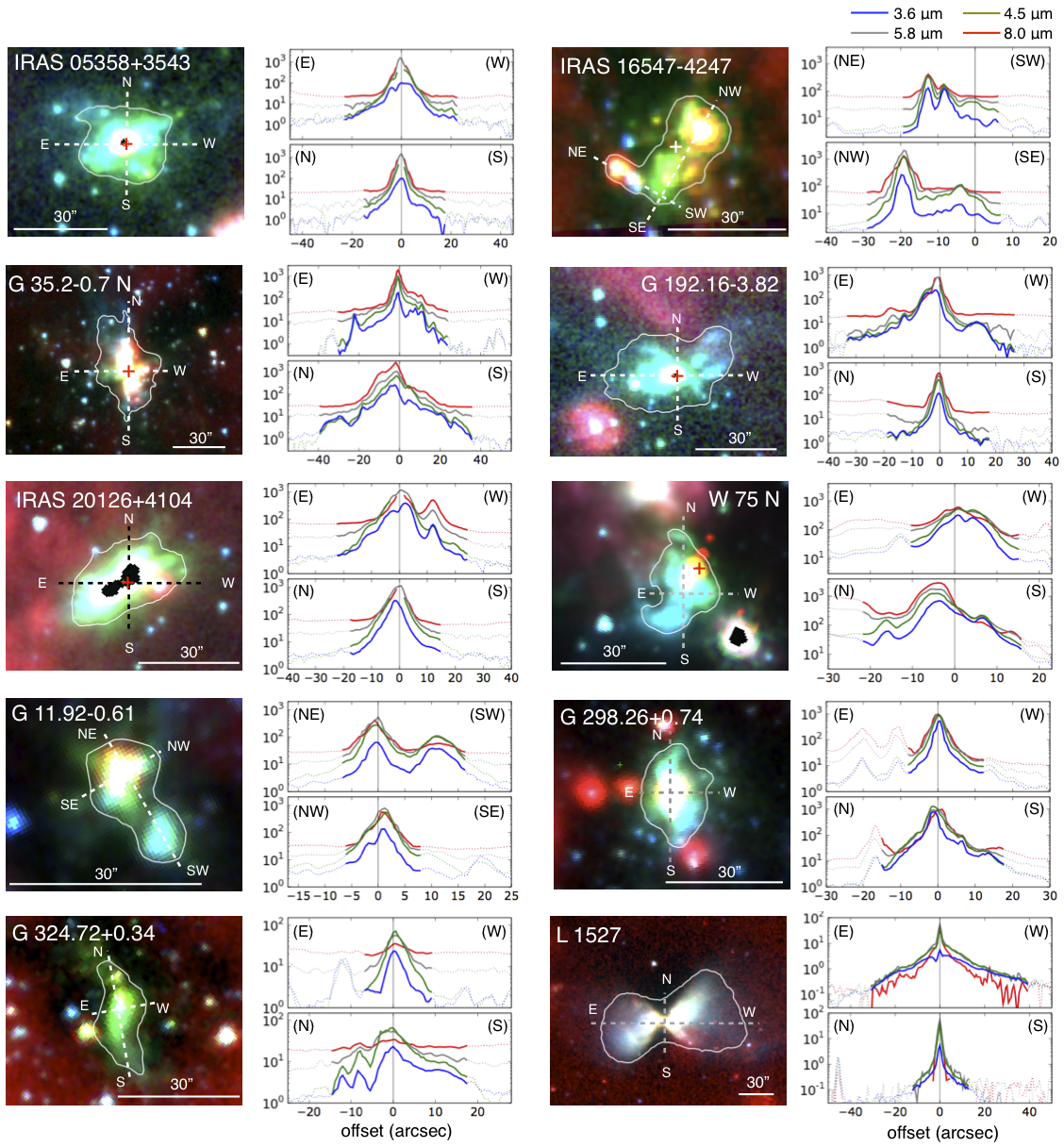}
%\plotone{f1.eps}
%\vspace{-0.7cm}
\caption{
Three-color images (blue, green, and red for 3.6, 4.5 and 8.0 \micron, respectively) and one dimensional intensity distributions (blue, green, gray and red for 3.6, 4.5, 5.8 and 8.0 \micron, respectively) for individual objects. {In all the three colors north is up and east is right.  For the these images} the color contrasts were set with the upper and lower limits based on the specified percentage of the flux in the entire image (see text).
%(99.5 \% for W 75 N; 98 \% for G 35.2--0.7 N, G 192.16--3.82, G 11.92--0.61, G298.26+0.74, G 324.72+0.34, and L 1527; 97 \% for IRAS 05358+3543, IRAS 20126+4104, and G 5.89--0.39 ; 96 \% for IRAS 16547--4247).
The saturated regions near these positions are masked in black.
The small cross in each three-color-image shows the position of the protostar (HPMOs, HC/UC H IIs) measured using millimeter interferometry (see Table \ref{target list} for references).
The white curve in the three-color image indicates the region where we measured the flux ratios and colors.
The positions where we extract the one dimensional intensity profiles are marked using dotted lines. In the one-dimensional profiles, the flux outside these positions is indicated using dotted curves. 
In these plots the flux is shown in logarithmic scale in MJy str$^{-1}$.
\label{fig_3color_1D}}
\end{figure*}

The Spitzer IRAC data in all four bands (3.6, 4.5, 5.8, and 8.0
\micron) were obtained from the Spitzer archive.  We use the pipeline
reduced post-BCD (Basic Calibration Data) images  provided by the
archive.  The mean FWHMs of the point response functions (PRFs) are
1".66, 1".72, 1".88 and 1".98, respectively.
%For accurate analysis
%with the flux ratios, we first
%carefully subtract the diffuse background emission using the
%techniques described in Takami et al. (2010).
%
%
Figure \ref{fig_3color_1D} shows
the conventional three color images: blue, green, red 
for 3.6, 4.5, and 8.0 \micron, respectively, as well as one-dimensional
cut-outs of the intensity distributions in all four bands. All the
objects are associated with Green Fuzzy emission, and
the one-dimensional profiles show that these have {a similar shape}
at 3.6, 4.5, and 5.8 \micron.

Our contrast adjustment for the three colors in Figure \ref{fig_3color_1D} was made
by setting the upper and lower limits based on the specified percentage ({95, 96, 97, 98, 99 or 99.5 \%},
depending on the objects) of the flux distribution in the larger areas. 
As a result, stars can appear ÒblueÓ due to the fact that
the flux is larger at shorter wavelengths; PAH emission can appear ÒredÓ due to
excessive emission at 8.0 \micron; the remaining extended emission can appear ÒgreenÓ
due to relatively low fluxes of stars and PAH at this wavelength.
While other authors \citep[][]{Noriega-Crespo04, Marston04, Rathborne05, SmithH06, Araya07, Kumar07, Shepherd07, Tobin07, Tobin08, Qiu08, Teixeira08, Cyganowski08, Cyganowski09, Cyganowski11, Chambers09, Morales09, Zhang09, Simpson09, Varricatt11} do not clearly state their color adjustments, we assume that most, if not all, of them made it in a more or less similar manner.
%For instance, \citet{Cyganowski08}, \citet{Cyganowski09}, and \citet{Cyganowski11} have provided images of G 35.2--0.7 N and G 11.92--0.61 with color contrasts similar to those in Figure \ref{fig_3color_1D}.
%For instance, the three-color image of G 11.92--0.61 in  \citet{Cyganowski08}, \citet{Cyganowski09}, and \citet{Cyganowski11} { also show a green color (but more extended due to different contrast adjustment)} as that in Figure \ref{fig_3color_1D}.

Despite the ``green'' color in the three color image,
none of the sources clearly show excess emission at 4.5 \micron~in the
one-dimensional cut-outs in Figure \ref{fig_3color_1D}.
Indeed, \citet{Chambers09}, who coined the term Green Fuzzies, define a quantitative criteria
for their detection as a 4.5 \micron~to 3.6 \micron~ ratio of $\ge$ 1.8 and a 5.8 \micron~to 4.5 \micron~ ratio of $\le$ 2.5.
Plotting this criteria, one can easily find that those with a marginal deficit at 4.5 \micron~can also be categorized
as Green Fuzzies. Throughout, the green color in three-color images may not always imply the presence of the 4.5-\micron~emission enhanced over the other IRAC bands as stated in the literature \citep[e.g.,][]{Rathborne05, Araya07, DeBuizer10}.

%Note that, the scale for three colors are somewhat arbitrary: these are automatically adjusted using ds9 developed by Smithonian Astronomical Observatoy {( True?)}. We believe the color contrast for many publications are made in the same manner (e.g. ....).
%While the green color of these images is due to emission at 4.5 \micron,  these extended emission components are visible at 3.6 \micron~ and 5.8 \micron~ as well.

Figure \ref{fig_3color_1D} also indicates the regions in each object where we apply our analysis with
flux ratio maps and the pixel by pixel color-color diagrams. {The criterion is described in Section 3.} Throughout the paper we focus the analysis on the brightest regions (contrast ratio to the brightest pixels to be 1/10--1/100) where we expect
high signal-to-noise.
%, and a significant fraction of the emission integrated over the space.
This means that our analysis misses the faint outer regions discussed in previous literature in the
W 75 N \citep{Qiu08} and G 11.92--0.61 regions \citep{Cyganowski08,Cyganowski09,Cyganowski11}\footnote{\citet{Cyganowski08} measured the total flux of G 11.92--0.61 of 0.334 Jy in the 4.5-\micron~band. In the area shown in Figure \ref{fig_3color_1D}, we measure the total flux of 0.257 Jy, implying that 77 \% of the flux is included.}.

\begin{figure*}
\epsscale{2.2}
%\epsscale{1.0}
\plotone{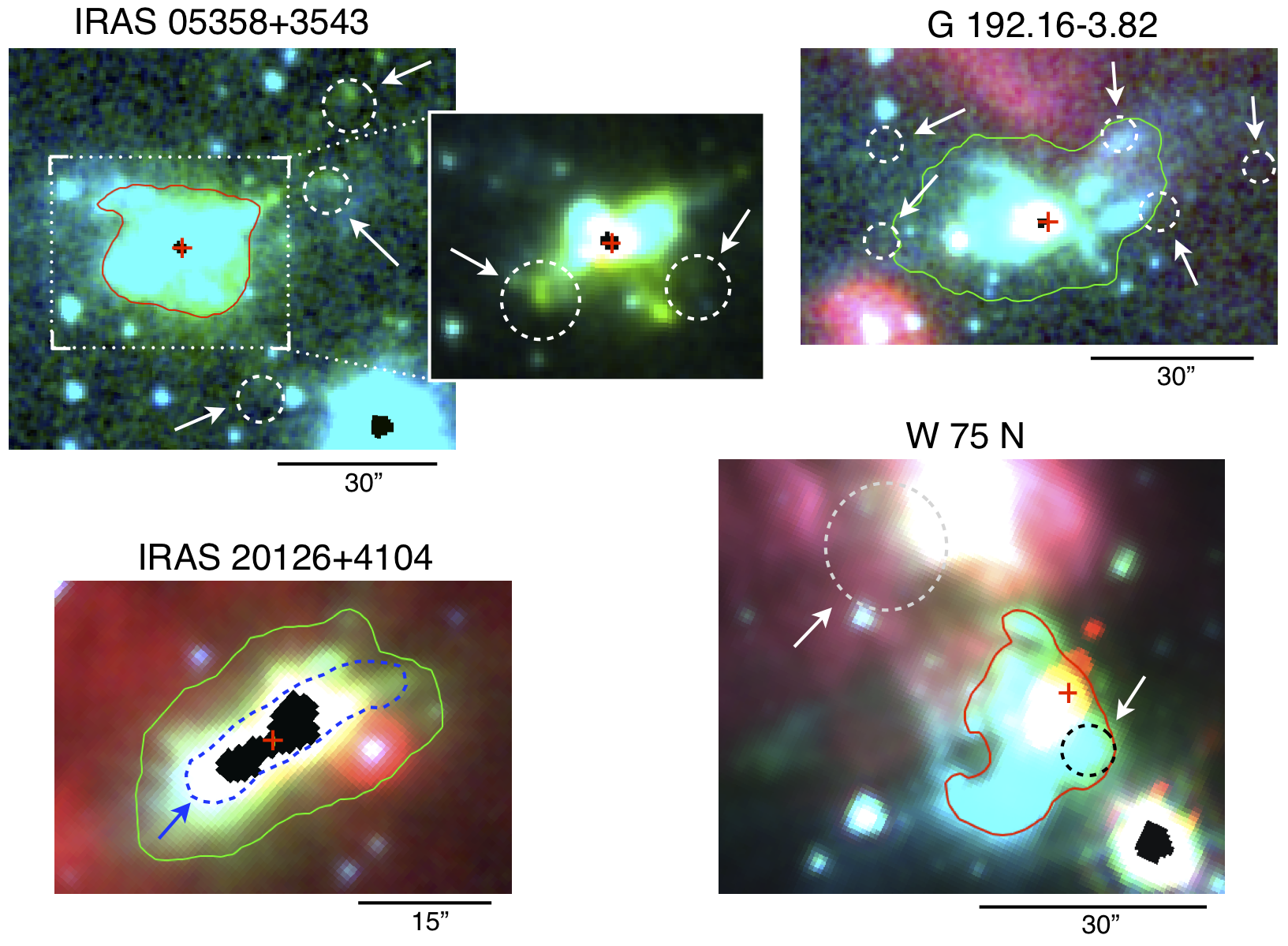}
%\plotone{f2.eps}
\caption{
Three-color images for IRAS 05358+3543, G 192.16--3.82, IRAS 20126+4104,
%G 5.89--0.39,
and W 75 N with different contrasts from Figure \ref{fig_3color_1D}.
{In all the three colors north is up and east is right.}
Dashed circles, curves and rectangles with arrow are the positions where the presence of H$_2$ 2.12 \micron~emission was reported in the literature (Varricatt et al. 2010 for IRAS 05358+3543, G 192.16--3.82, and IRAS 20126+4104; 
%Puga et al. 2006 for G 5.89--0.39;
Davis et al. 1998 for W 75 N).
The regions where we measured the flux ratios and colors are also indicated using red or green curves.
%The letters A, B, C in the G 5.89--0.39 are identifications by \citet{Puga06}. The red emission in and at the upper side of C is an artifact.
\label{fig_IRAC_vs_obsH2}}
\end{figure*}

Imaging observations of H$_2$ at 2.12 \micron~ 
%and $K$-band continuum
have been made near these bright regions for all the nearby high-mass protostar sample. These are:-
IRAS 05358+3543 \citep{Varricatt10},
IRAS 16547--4247 \citep{Brooks03},
G 35.2--0.7 N \citep{Froebrich11,Lee12},
G 192.16--3.82 \citep{Indebetouw03, Varricatt10},
IRAS 20126+4104 \citep{Shepherd00,Varricatt10}, 
%G 5.89--0.39 \citep{Puga06},
and W 75 N \citep{Davis98a,Shepherd03}.
%, and
%G 11.92--0.61 (Lee et al., in preparation).
%The H$_2$ 2.12 \micron~emission in IRAS 05358+3543, G 192.16--3.82, G 5.89--0.39, and W 75 N are identified as knots, while that in IRAS 20126+4104 exhibits a collimated jet. 
The H$_2$ 2.12 \micron~emission in IRAS 05358+3543, G 192.16--3.82, and W 75 N are identified as knots, while that in IRAS 20126+4104 exhibits a collimated jet. 
These contrast to the fuzzy morphology of the green emission in Figure \ref{fig_3color_1D}. Figure \ref{fig_IRAC_vs_obsH2} shows the locations of the H$_2$ emission in the three-color images. At these positions only faint 4.5-\micron~emission components are seen at some of these positions. The H$_2$ 2.12 \micron~emission in IRAS 16547--4247 and G 35.2--0.7 N show more complicated structures in shocked gas  (Brooks et al. 2003; Froebrich et al. 2011, Lee et al. 2012). {These have a significantly larger spatial extension (by a factor of 2--3) than the bright IRAC emission shown in Figure \ref{fig_3color_1D}, and most of the flux is distributed outside the region indicated}.
%
% Froebrich et al. (2011) ... K
% Lee et al. (2012) ... K
% Varricatt et al. (2010) ... K
% Brooks et al. (2003) ... no continuum
% Indebetouw et al. (2003) ... K'
% Shepherd et al. (2000) ... Ks
% Puga et al. (2006) ... Ks
% Davis et al. (1998) ... 2.140 micron
% Shepherd et al. (2003) ... Ks
%   G 35, G 192  ... similar bipolar geometry both in IRAC and K-band images
%   G 192 ... a horn-like structure
%   I05, both elongated in the EW, but its is much more compact in K-band
%   W 75... both extended in the south-east direction
%
%

Most of the above literature also provide the continuum images at $\sim$2 \micron, either $K (2.2 \micron)$, $K$' (2.12 \micron), $K_s$ (2.15 \micron) or 2.14 \micron. Their distributions in IRAS 05358+3543, G 35.2--0.7 N, G 192.16--3.82, IRAS 20126+4104, and W 75 N are remarkably different from H$_2$ 2.12 \micron, but to some extent similar to the IRAC emission in Figure \ref{fig_3color_1D}. In particular, a bipolar structure is observed in G 35.2--0.7 N and G 192.16--3.82 in both IRAC and {the continuum at 2-\micron} \citep[see][]{Froebrich11,Indebetouw03,Varricatt10}. In the west of G 192.16--3.82 both IRAC emission and {2-\micron~continuum} show a horn-like structure at similar angular scales. In W 75 N, both IRAC emission and 
%$K$-band
{2-\micron~Œcontinuum} are extended to the south-east side of the protostar, in contrast to the orientation of the molecular outflow \citep{Shepherd03} and the distribution of H$_2$ emission in different directions \citep{Davis98a, Shepherd03}.
%In G 5.89--0.39, \citet{Feldt03} and \citet{Puga06} provide the 2-$\micron$~continuum image only in the close proximity to the protostar, and its emission distribution approximately coincides with the saturated area in our IRAC data.
The detection of the {2-$\micron$ continuum} at these positions indicates that the different distribution seen in the bright parts of Green Fuzzy emission and H$_2$ 2.12 \micron~may not be attributable to extinction \citep{Lee12}. 

% IRAS 05358+3543: collimated jet in molecular emission (Qiu et al. 2008)
% IRAS 16547: 
% 
% 
% 
% 
% 
Molecular outflows have been observed in most of our nearby high-mass protostar sample. G 35.2--0.7 N and G 192.16--3.82 are associated with molecular outflows with similar morphology to the direction of elongation in the Green Fuzzy emission \citep{Gibb03,Shepherd98}, but the position angles between the Green Fuzzies and molecular outflows are different by $\sim 30^{\circ}$. In IRAS 05358+3543
%, G 5.89--0.39,
and W 75 N, the distribution of Green Fuzzy emission is remarkably different from the molecular outflows observed at millimeter wavelengths by
%\citet{Beuther02, Hunter08,Davis98a, Shepherd03}
\citet{Beuther02, Davis98a, Shepherd03}
\citep[see also][]{Qiu08}.
For IRAS 16547--4247, \citet{Brooks03} shows the presence of extended shocked H$_2$ emission to the north and south of the protostar. This contrasts to the distribution of Green Fuzzy emission in Figure \ref{fig_3color_1D}.
In IRAS 20126+4104 \citet{Cesaroni99} shows the presence of a molecular outflow in the millimeter HCO$^+$ and SiO emission with the same orientation as the Green Fuzzy emission in Figure \ref{fig_3color_1D}, but with a slightly smaller angular scale.
%It is intriguing that 
This object is also associated with a more extended outflow in the millimeter CO emission with a remarkably different orientation \citep{Shepherd00}.

%The southwest peak of G 11.92-0.6 is associated with H$_2$ 2.12-\micron~emission and $K$-band continuum (Lee et al., in preparation), and also a molecular outflow in millimeter emission \citep{Cyganowski11}. The bright peak in H$_2$ emission is offset by $\sim$3" to the southwest from the peak in the IRAC emission, and its fainter component is elongated in the northeast to the southwest direction, analogous to the blueshifted lobe of the molecular outflow observed by \citet{Cyganowski11}. This contrasts with the $K$-band continuum, whose distribution approximately matches the IRAC emission. The $K$-band continuum is also observed at the northeast peak of the IRAC emission with a lower flux.

\section{Flux Ratio Maps and Color-Color Diagrams}

\begin{figure*}
\epsscale{2.2}
%\epsscale{1.0}
\plotone{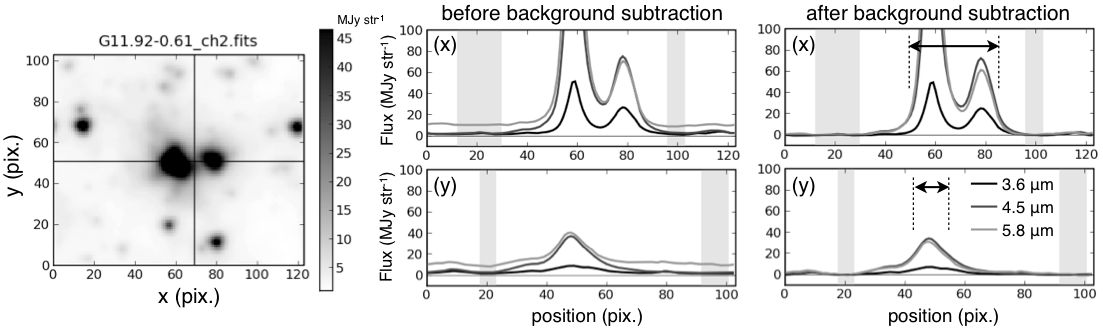}
%\plotone{f3.eps}
\caption{
Examples of background subtraction. ($left$) the $I_{4.5\micron}$ image of the objects {after convolution}. The coordinates are shown in pixels (1 pixel correspond to 0".6). The crosshair in the image shows the positions where we measured the one-dimensional flux distribution for background subtraction. ($middle$) one-dimensional intensity profiles at 3.6, 4.5 and 5.8 \micron~ before subtracting the background. Gray squares show the range where we measured the background level and its standard deviation after background subtraction (i.e., the uncertainty of the flux measurement for the flux ratio maps and color-color diagrams). ($right$) same as the middle but after background subtraction. {The arrows and dashed lines show the spatial ranges for $>$$15 \sigma$ ($>$18.7 MJy str$^{-1}$) for the 4.5-\micron~emission. This corresponds to the region we selected for the color-color diagrams (see text for details).}
\label{fig_bg_subtraction}}
\end{figure*}

In most of the regions the emission at 8.0 \micron~ is severely contaminated by
bright diffuse extended emission not directly associated with the
protostar. Therefore, we will use the data for 3.6, 4.5, and
5.8 \micron, excluding 8.0 \micron.
%To produce accurate flux ratio maps and color-color diagrams, the individual images need to be carefully background subtracted and matched in resolution. 
To produce accurate flux ratio maps and color-color diagrams, the individual images need to be matched in resolution and carefully
background subtracted.
We first convolve each image with the appropriate PRFs to try to cancel out the effect of different
PRFs at 5.8 \micron~from 3.6 and 4.5 \micron~(note that the PRFs at 3.6 and 4.5 \micron~
are almost identical --- see IRAC Data Handbook 3.0). 
The 3.6 and 4.5 \micron~images are convolved with the PRF at 5.8
\micron, and the 5.8 \micron~images with the PRF at 3.6 \micron. 
This yields an effective angular resolution of $\sim$3".
We then subtract the diffuse background emission by measuring
it in the $x$- and $y$-directions, and fitting it using a linear function.
Figure \ref{fig_bg_subtraction} shows how this process works for {one} of our sample.
{This fitting process simultaneously allows us to measure the uncertainty of flux, which is dominated by
non-uniform distribution of diffuse emission. These are tabulated in Table \ref{target list}.}

\begin{figure*}
\epsscale{2.2}
%\epsscale{1.0}
\vspace{-2.5cm}
\plotone{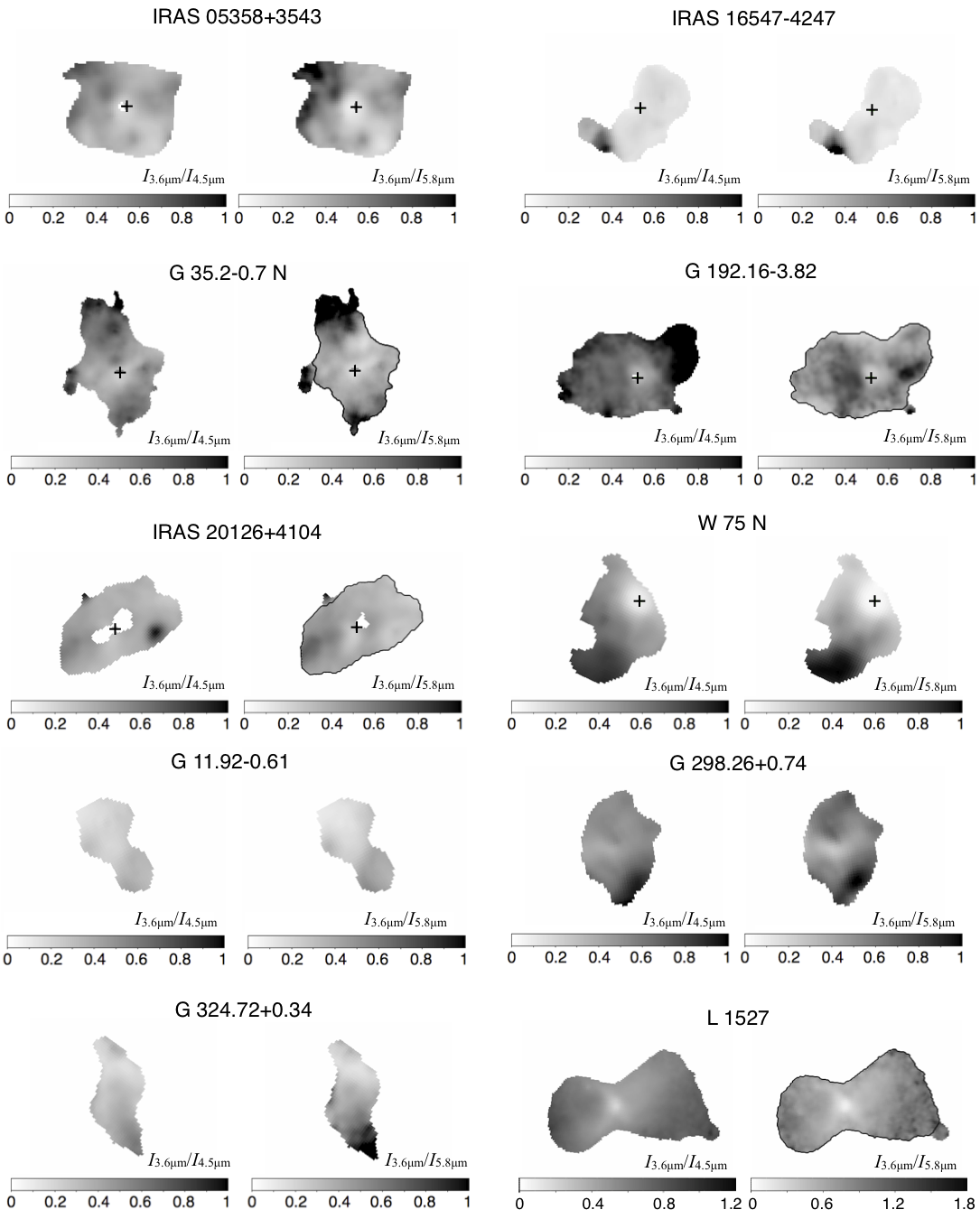}
%\plotone{f4.eps}
\figcaption{
The $I_{3.6 \micron}/I_{4.5 \micron}$ and $I_{3.6 \micron}/I_{5.8 \micron}$ maps for individual objects.
{ In all the three colors north is up and east is right.}
The region is selected based on the criterion in Table \ref{target list}.
% for the regions shown in Figure \ref{fig_3color_1D}. 
The range of grayscale for the $I_{3.6 \micron}/I_{4.5 \micron}$ and $I_{3.6 \micron}/I_{5.8 \micron}$ maps is 0--1 for all figures but L 1527 (0--1.2 and 0--1.8 for this object).  The cross shows the position of the protostar (HPMOs, HC/UC H IIs) measured using millimeter interferometry  (see Table \ref{target list} for references). The saturated regions near these positions are masked in white. Black curves are drawn in the $I_{3.6 \micron}/I_{5.8 \micron}$ maps for G 35.2-0.7 N, G 192.16-3.82, IRAS 2016+4104 and L 1527, and these are the regions selected for color-color diagrams excluding stars near the boundary. For the other objects the entire region shown in the figure is used for color-color diagrams.
\label{fig_flux_ratios}}
\end{figure*}

To investigate the origin of the Green Fuzzy emission, we use pixel by pixel color-color diagrams
with [3.6]-[4.5] and [3.6]-[5.8]. {As described in Section 2, we focus the analysis on the brightest regions where we expect the high signal-to-noise to clearly show the tendencies in color-color diagrams described later. The criterion
for selection is summarized in Table 2. The selection is made  based on the 4.5-\micron~flux: $>$15-$\sigma$
for most of the objects. For W 75 N and G 298.26+0.74 a larger limit is applied to exclude the emission associated
with other object(s) nearby. For IRAS 20126+4104 we apply an additional flux limit ($>$5-$\sigma$) for 5.8-\micron.
This allows us to remove artifacts in the color-color diagrams that are due to a high uncertainty in the background subtraction at this
wavelength for this object. For L 1527 a flux limit of $>$25-$\sigma$ is applied to clearly show the linear correlation between 
[3.6]-[4.5] and [3.6]-[5.8] described later. In all the objects we rebin the data onto a 3'' grid, i.e., the same angular scale as
the resolution after convolution.
%The individual images are carefully
%background subtracted and the effect of different PRF are corrected using the same methods
%described above. 
%After the convolution and background subtraction described above, we take the additional step of masking regions with low signal to noise and saturated regions (Figure \ref{fig_3color_1D}). 
%Finally, we rebin the data onto a 3'' grid to improve the signal-to-noise ratio.

Figure \ref{fig_flux_ratios} shows the 3.6/4.5- and 3.6/5.8-\micron~ flux ratios maps in the regions described above.
This figure shows that the the distributions of the 3.6/4.5- and 3.6/5.8-\micron~ flux ratios
are similar for each object except G 192.16--3.82.
IRAS 20126+4104, G 35.2-0.7N, G192.16-3.82, and L 1527 are associated with stars near the boundary. These appear blue in three-color images in Figure \ref{fig_3color_1D}, and exhibit high 3.6/4.5- and 3.6/5.8-\micron~ flux ratios ratios in the flux ratio maps in Figure \ref{fig_flux_ratios}. The color-color diagrams for these objects are made excluding these stars. The regions after this removal are indicated in Figure \ref{fig_flux_ratios} as well as Figure \ref{fig_3color_1D}. In IRAS 20126+4104, the 3.6/4.5-\micron~ flux ratio map shows another point source with high
flux ratios in the west of the region. This source is absent or marginal in the 3.6/5.8-\micron~map. The inferred color of this source and the rest of the region is discussed later.

In Figure \ref{fig_flux_ratios} we also plot the positions
of the protostar (HPMO, HC/UC H II) measured using millimeter interferometry. Although
the position is in the mask of low signal-to-noise or a saturated region for a few objects,  this approximately matches
the region with the lowest 3.6/4.5- and 3.6/5.8-\micron~flux ratios. Similarly, these flux ratios are the lowest  at the center
of L 1527 where the protostar is located \citep{Tobin08}.

%Figure \ref{fig2} shows the color-color diagram
%for L 1527, the reference for the scattered continuum associated with low-mass protostars,
%with the colors of shocked emission measured in several Herbig-Haro objects  \citep[][]{Takami10b},
%PAH emission modeled by \citet[][]{DL07} for 
%a wide range of UV fields and PAH abundance  ($U$=1--$1 \times 10^5$ and $q$=0.47--4.58, where $U$ is the UV field normalized to the interstellar radiation field,  $q$ is the mass fraction of PAH in percent),
%and PAH emission observed in the reflection nebula NGC 7023 (Spitzer IRAC Data Handbook).
%In Figure \ref{fig2} the [3.6]-[4.5] colors for each 3"$\times$3" image bin 
%measured in L 1527 show a remarkable correlation
%along the extinction vector measured in molecular clouds \citep{Chapman09}. The result can be fitted with
%a regression line,  ({[3.6-[5.8])=([3.6]-[4.5])$\times$1.405--0.097. In Figure \ref{fig2} the shocked emission is distributed
%at both sides of this regression line like an arc, at [3.6]-[4.5] and [3.6]-[5.8] of 1.1--2.3 and 1.4--3.5, respectively.
%The colors measured in L 1527 and shocks overlap in a relatively small color range: at [3.6]-[4.5] and [3.6]-[5.8] of 1.2--1.6
%and 1.7--2.1, respectively.
%PAH emission is characterized by a combination of relativelly small [3.6]-[4.5] and large [3.6]-[5.8] (--0.6 to 0.3 and 2.4 to 4.0, %respectively).

Figure  \ref{fig_cc_hmonly}
shows the colors measured in the 3$\times$3 arcsec image
bins for Green Fuzzy emission in each high-mass star forming region.
%The [3.6]-[4.5] vs [3.5]-[5.8] colors in seven objects (IRAS 05358+3543, IRAS 16547--4247, G 35.2--0.7 N, W 75 N, G 11.92--0.61, G298.26+0.74 and G 324.72+0.34) show a fairly strong correlation along the direction of the extinction vector.
All the objects but IRAS 20126+4104 and G 192.16-3.82 show a fairly strong correlation between the [3.6]-[4.5] and [3.6]-[5.8] colors along the direction of the extinction vector.
Their values vary between objects in the ranges of [3.6]-[4.5] and [3.6]-[5.8] of 0.4--3 and 0.1--4.7, respectively.
%A similar trend is also observed in a majority of pixels selected in Green Fuzzy emission associated with two HC HII regions G 192.16--3.82 and IRAS 20126+4104. Colors in some positions in G192.16--3.82 show a [3.6]-[5.8] excess from this correlation peaking at [3.6]-[4.5] $\sim$0.5. In IRAS 20126+4104, most of the region shows a good correlation between [3.6]-[4.5] and [3.6]-[5.8], while [3.6]-[5.8] remains relatively constant ($\sim$2.4) over [3.6]-[4.5]=0.7--1.9 at several positions.
%G5.89--0.39 does not clearly show a correlation
%between [3.6]-[4.5] and [3.6]-[5.8] at any color range.
A similar trend is also observed in most of pixels selected in Green Fuzzy emission associated with {IRAS 20126+4104. In addition to the component in which the [3.6]-[4.5] and [3.6]-[5.8] colors show a linear correlation, this object is associated with several pixels for which [3.6]-[5.8] is constant (2.2--2.4) over [3.6]-[4.5] =0.8--1.5.
Some pixels in G 192.16-3.82 also show a linear correlation like that described above, however, this object is also associated with a number of pixels with larger [3.6]-[4.5] and [3.6]-[5.8] colors, ranging between 0.2--1.7 and 1.0--2.4 for 
[3.6]-[4.5] and [3.6]-[5.8], respectively.
}
%Colors in some positions in G192.16--3.82 show a [3.6]-[5.8] excess from this correlation peaking at [3.6]-[4.5] $\sim$0.5. In IRAS 20126+4104, most of the region shows a good correlation between [3.6]-[4.5] and [3.6]-[5.8], while [3.6]-[5.8] remains relatively constant ($\sim$2.4) over [3.6]-[4.5]=0.7--1.9 at several positions.

Figure \ref{fig_deviation1} shows the locations of the regions which have a [3.6]-[5.8] excess from the linear correlation in G 192.16--3.82, and where the [3.6]-[5.8] color has a constant value in IRAS 20126+4104. In G 192.16--3.82 the [3.6]-[5.8] excess is located in the outer regions where the 4.5-\micron~flux is relatively faint. In IRAS 20126+4104 the emission with a constant [3.6]-[5.8] is located around a point source to the west of the protostar. The latter is also clearly seen in the 3.6/4.5-\micron~flux ratio map in Figure \ref{fig_flux_ratios}. These areas are responsible for a small fraction of the 4.5-\micron~flux as compared with that of the entire area shown with dashed curves: these are 22 \% and 6 \% for G 192.16--3.82 and IRAS 20126+4104, respectively.

Some authors mention the possibility that such ``green emission" is due to CO ro-vibrational transitions at 4.5-5 \micron~\citep[see e.g.,][; see also Takami et al. 2010 and reference therein]{Shepherd07,Cyganowski08, Cyganowski09,Chambers09}. However, our results do not show clear evidence for such emission components, which should only appear in the 4.5-\micron~band. 
If the Green Fuzzy emission in these objects is associated with shocked emission as discussed by some authors (see Section 1), the absence of CO emission contrasts to the shocks associated with low-mass protostars, which also show green colors in three-color images \citep[e.g.,][]{Noriega-Crespo04,Tobin07,Teixeira08,Zhang09}. In the latter a 4.5-\micron~excess, presumably due to CO emission, is observed at some positions in most of the objects in Takami et al. (2010, 2011). In the next two sections we discuss the emission mechanism of Green Fuzzy emission associated with high-mass protostars and their candidates.}
%The main emission component we show above has a good correlation between the [3.4]-[4.5] and [3.5]-[5.8] colors. All the other emission components show smaller [3.6]-[4.5] colors, implying that the 4.5-\micron~emission is less strong. The absence of the signature for CO emission is consistent with the following fact. Thermal excitation of CO ro-vibrational transitions requires the gas density to be $> 10^8$ cm$^{-3}$ to dominate over thermal shocked H$_2$ emission \citep[][]{Neufeld08}, which appears in all four IRAC bands \citep[e.g.,][]{Smith05a, Takami10b}. Such a density is too high to be expected in normal molecular clouds \citep[$< 10^4$ cm$^{-3}$,][]{Stahler05}.

\section{Comparisons with Other Objects and Models in Color-Color Diagrams}

In the following subsections we compare these colors with
scattered continuum (Section 4.1),
molecular shocks associated with low-mass protostars and modeled thermal H$_2$ emission (Section 4.2),
and PAH and fluorescent H$_2$ emission (Section 4.3).

\subsection{Scattered continuum}
\begin{figure*}
\epsscale{2.2}
%\epsscale{1.0}
\plotone{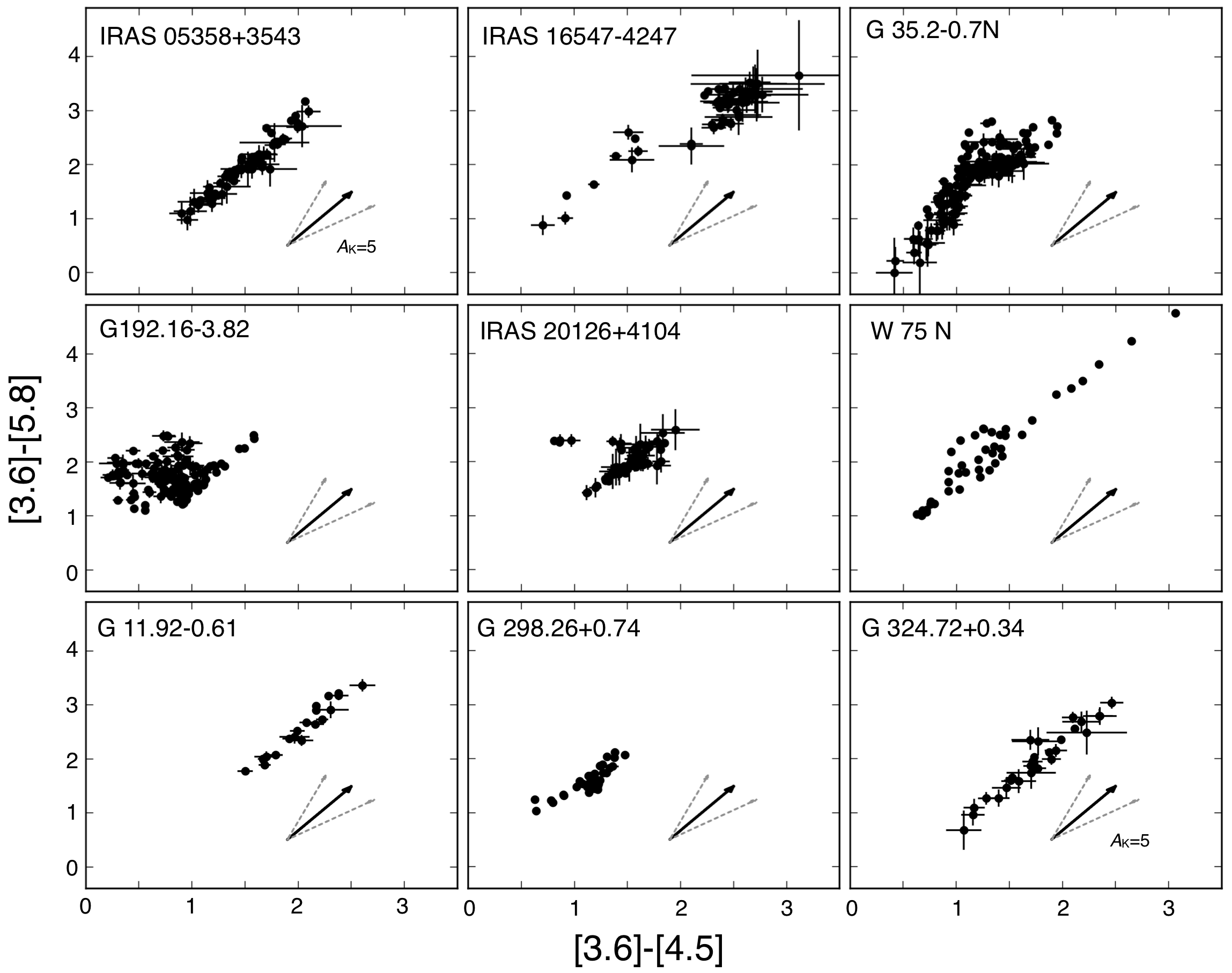}
%\plotone{f5.eps}
\caption{
The [3.6]-[4.5] versus [3.6]-[5.8] colors measured in ``Green Fuzzies'' in ten high-mass star forming regions. The error bars are shown only for those larger than the size of the dots. The solid arrows show the extinction vector based on the measurements of molecular clouds ([3.6]-[4.5] and [3.6]-[5.8] of 0.09$\pm$0.03 and 0.18$\pm$0.03, respectively, at $A_K \ge 2$; Chapman et al. 2009). Gray arrows show the extinction vectors at $\pm$1-$\sigma$.
\label{fig_cc_hmonly}}
\end{figure*}

Figure  \ref{fig_cc_with_L1527} shows the same color-color diagrams as Figure  \ref{fig_cc_hmonly}
%\ref{fig_cc_hmonly}
but we include  the points for scattered continuum from the L 1527 protostellar outflow in each diagram.
%Figure \ref{fig3} shows the colors measured in Green Fuzzy emission, scattered continuum in L 1527,
%and shocked emission.
The correlation of [3.6]-[4.5] and [3.5]-[5.8] colors observed in most of the high-mass star forming regions is remarkably similar to that of L 1527. The regression line for L 1527 fits  the observed colors in IRAS 05358+3543, IRAS 16547--4247, G 298.26+0.74, and most of the colors
measured in IRAS 20126+4104 as well. The figure shows a slightly larger [3.6]-[5.8] color
% (by  $\sim$0.5)  in G 35.2--0.7 N
{(up to 0.6)  in W 75 N}
and a slightly smaller [3.6]-[5.8] color (by {0.3--0.7}) in G 11.92--0.61 and G 324.72+0.34 compared with the regression line.  G 35.2--0.7 N and W 75 N show a [3.6]-[5.8] excess at [3.6]-[4.5]=1.0--1.5.

If the Green Fuzzy emission in our sample of high-mass star forming regions is due to scattered
continuum, the observed colors should be a function of: (1) the
intrinsic color of the central star(+disk); (2) the color change due to
scattering; and (3) extinction. To investigate the first two issues,
we have made simplified calculations with existing star-disk models
and dust models. The modeled colors for star-disk systems were obtained
from \citet{Robitaille06}, based on radiative transfer
calculations for protostellar environments with the protostar, disk,
envelope and outflow cavities by \citet{Whitney03a}. \citet{Robitaille06}
 present a grid of infrared spectral energy distributions
(SEDs) for 200,000 cases, with a variety of stellar and disk
parameters, at ten viewing angles for each model. To derive the
infrared flux from the disk, these authors include ``activeÓ
viscous heating based on the standard accretion disk model by \citet{Shakura73}, and ``passiveÓ heating by stellar
radiation. They also adopt the disk geometry of \citet{Shakura73}.

To derive the intrinsic color of the star-disk systems {(i.e., ``1" described above)}, we select all the
modeled results with $A_V < 5$ (corresponding to
$E_{3.6\micron-4.5\micron} < 0.07$ and $E_{3.6\micron-5.8\micron} <
0.12$ for the adopted grain model). {This criterion is applied due to the fact that
the \citet{Robitaille06} models apparently include dust in the envelope
and outflow cavity so that we cannot derive the values without extinction.
The above criterion allows us to include a large number of samples with a variety
of disk masses and accretion rate with small errors in color due to extinction.}
We also select all the results
with a stellar mass larger than 8 $M_\odot$. As a result, a total of
17,580 SEDs are selected from the grid, with a range of stellar mass
$M_*$=8--50 $M_\odot$, stellar effective temperature $T \sim 4 \times
10^3 - 5 \times10^4$ K, disk mass 0--3 $M_\odot$, and accretion rate
$M \sim 10^{-14}-10^{-3}~M_\odot$ yr$^{-1}$ if the system hosts a
disk.  In addition to the modeled colors of \citet{Robitaille06},
we calculate the color for blackbodies with a single temperature of
$T$=500, 1000 and 4000 K.

The color change via single scattering is derived from \citet{Robitaille06}, with the grain models based on the optical constants and size distribution by \citet{Laor93} and \citet{Kim94}, respectively. We consider the following two cases for scattering: (1) scattering in optically thin regions, i.e., where the color change via scattering is determined by the scattering cross section at each wavelength; and (2) scattering for optically thick and geometrically thin cavity walls, which may be realistic for outflow cavities associated with low-mass protostars \citep[e.g.,][]{Whitney03b, Tobin08}. For the latter, the efficiency of the scattering should be approximately determined by the scattering albedo, assuming that the effects of internal extinction and multiple scattering are negligible. For the color change due to extinction, the extinction vector based on \citet{Chapman09} has a large uncertainty as shown in Figure \ref{fig_cc_hmonly}. Here we assume that the linear correlation observed in L 1527 is due to different extinction at different positions, and extract the values from the regression line.

Figure \ref{fig_cc_with_scamodels} shows the modeled colors of the star-disk systems with
scattering together with the observed colors. This figure shows that
colors of these objects approximately lie within the extinction track of either
the optically thin or thick results. In this case, the linear correlation of the [3.6]-[4.5] and
[3.6]-[5.8] colors in each object is attributed to different extinctions at different positions.
%This implies that, in this case, the 3.6/4.5-\micron~and 3.6/5.8-\micron~maps represent the distribution of extinction.
The different objects show different colors not only
in the direction of the extinction vector, but also across it. The latter is consistent with the idea that
the intrinsic color of the star(+disk) systems differs between objects. In IRAS 05358+3543, G 11.92--0.61, and G 324.72+0.34, the
intrinsic flux of the central source is dominated by a star with
a relatively high color temperature ($T >$1000 K). In W 75 N and G
298.26+0.76, the central source is redder, i.e., its flux is
dominated by a protostar with a relatively low color temperature ($T
<$1000 K).
%Follow-up spectroscopy would let us investigate this
%issue in detail,  allowing us to discuss the presence or
%absence of an inner disk, which is useful for testing the formation
%scenario of high-mass protostars (Beuther et al. 2007; Cesaroni et
%al. 2007). 
Figure \ref{fig_cc_with_scamodels} also shows that the two different scattering models
yield similar results for the color of the embedded central
source. 
%If we were able to identify the condition of the dust (i.e.,
%either optically thin, or optically thick and geometrically thin), this
%diagram would also allow us to measure the absolute value of
%extinction towards these objects.

In G 35.2--0.7 and W 75 N the distribution of the correlation is broader, the [3.6]-[5.8] color partially exceeding those expected for the models
at [3.6]-[4.5] $\sim$1. Figure \ref{fig_deviation2} shows their distribution in the 4.5-\micron~map. These areas are responsible for 34 \% and 9 \% of the
4.5-\micron~flux integrated over the entire area shown in dashed curves in G 35.2--0.7 and W 75 N, respectively.

\subsection{Shocks in low-mass protostellar jets, thermal H$_2$ emission}

Figure \ref{fig_cc_HH} shows the same color-color diagrams as Figures \ref{fig_cc_hmonly} and \ref{fig_cc_with_L1527} but with the colors observed in six  low-mass protostellar jets analyzed by \citet{Takami10b}. They showed that while some of them are explained by thermal H$_2$ emission (a combination of rotational and ro-vibrational transitions), others requires additional emission at 4.5 \micron, presumably due to the vibrationally excited CO emission. In Figure \ref{fig_cc_HH}, the colors are distributed along an arc with [3.6]-[4.5] and [3.6]-[5.8] of 1.1--2.3 and 1.4--3.5, respectively, located at both sides of the linear color correlation observed in most of the objects. While the observed correlations between [3.6]-[4.5] and [3.6]-[5.8] are totally different between Green Fuzzy emission and shocks associated with low-mass protostars,
their colors match in some color ranges in most of the objects. The range of such colors are  [3.6]-[4.5] = 1.2--1.6 and [3.6]-[5.8] = 1.5--2.5 for IRAS 05358+3543, G 192.16--3.82, IRAS 20126+4104, W 75 N and G298.26+0.74. This color range is slightly larger in G 35.2--0.7 N ([3.6]-[4.5] = 1.1--1.6 and [3.6]-[5.8] = 1.6--{2.8}), while IRAS 20126+4104 has an additional overlap at [3.6]-[4.5] = {1.3} and [3.6]-[5.8] = {2.4}.
%In G 5.89--0.39, the color range of the overlap with shocks is seen at [3.6]-[4.5] = 1.8 and [3.6]-[5.8] = 3.3, i.e., a different color range from the other, because of its markedly different distribution in the color-color diagram.

Figure \ref{fig_cc_H2_models} shows the colors for thermal H$_2$, i.e., the primary source of H$_2$ emission in shocks, based on calculations by \citet{Takami10b}. As in \citet{Takami10b} the models are made for iso-thermal cases and shock slabs with a power-law cooling function ($\Lambda \propto T^{- \alpha}$). The latter has been used to explain shocked H$_2$ emission observed both from the ground \citep[e.g.,][]{Brand88, Gredel94, Everett95, Richter95, Takami06a} and in space \citep{Neufeld08,Neufeld09,Takami10b}. For calculations of non-local-thermal-equilibrium (non-LTE)  cases thermal collisions are included for two cases, H+He and H$_2$+He, corresponding to gas with relatively high ($\gg 0.002-0.02$) and low ($\ll 0.002-0.02$) dissociation rates \citep[][]{Takami10b}. We adopt  $A$-coeffcients provided by \citet{Wolniewicz98}, and collisional rate coefficients for H$_2$ and He by \citet{LeBourlot99}. For collisional rate coefficients of H, we adopt \citet{Wrathmall07} and \citet{LeBourlot99}, and show the results separately. According to \citet{Wrathmall07}, they provide rate coefficients with a better accuracy than \citet{LeBourlot99} due to improved representation of the vibration eigen functions. In contrast, the coefficients provided by \citet{LeBourlot99} can explain the observed IRAC colors better in \citet[][]{Takami10b}.

The energy levels we include for calculations are 245 ($E_u/k=0-43000$ K) for LTE based on all the levels included in \citet[][]{DB96}; the lowest 49 energy levels (46 lines in the IRAC four bands, $E/k$ up to $2 \times 10^4$ K) for non-LTE H$_2$ with H+He collisions; and the lowest 36 energy levels (32 lines in the IRAC four bands, $E/k$ up to $1.69 \times 10^4$ K) for non-LTE H$_2$ with H$_2$+He collisions. The number of transitions included for non-LTE calculations is limited by the availability of collisional rate coefficients. The level populations of ortho- and para-H$_2$ are calculated separately, and those fluxes are combined assuming an ortho/para ratio of 3. The upper limit of temperature is set at 4000 K, approximately corresponding to the dissociation temperature of H$_2$ molecules \citep[e.g.,][]{Lepp83}. We set the lowest temperature of 30 K for numerical integration of the temperature slabs for power-law cooling. Since the emission in IRAC bands should originate from gas at much higher temperature \citep[$\ga$1000 K][]{Takami10b}, this limit does not affect the colors we show in Figure \ref{fig_cc_H2_models}. {For all the above cases the measured spectral response functions are used to obtain the IRAC fluxes.}

%\\ \\
%{ (... this paragraph is split into two ...)}
%\\ \\
Due to the limited number of transitions included, the accuracy of non-LTE calculations are low for the IRAC 3.6-\micron~band at the highest temperatures. The gaps between such calculations and those with 245 levels  are also shown in Figure \ref{fig_cc_H2_models} for the LTE regimes. The non-LTE gas of our calculations does not show smaller [3.6]-[4.5] and [3.6]-[5.8] colors than LTE, since the 3.6-\micron~emission requires a higher density for thermalization than the emission in the other IRAC bands \citep{Takami10b}.
%It would be filled if the collisional rate coefficients were available for transitions at high energy levels.
%Although our calculations are not complete on this point, it is not likely the non-LTE H$_2$ has smaller [3.6]-[4.5] and [3.6]-[5.8] colors since the 3.6-\micron~requires a higher density for thermalization than the emission in the other IRAC bands \citep{Takami10b}.

In Figure \ref{fig_cc_obs_vs_H2} we plot the observed colors in Green Fuzzy emission with possible colors due to thermal H$_2$ as shown in Figure \ref{fig_cc_H2_models}. While thermal H$_2$ could explain the colors observed at most of the positions, it cannot explain the colors with small [3.6]-[4.5] and [3.6]-[5.8] colors ($<$1.0 and $<$1.5, respectively) observed in {IRAS 05358+3543,} IRAS 16547--4247, G 35.2--0.7 N, G 192.16--3.82, W 75 N and G 298.26+0.74. 
{These correspond to the regions with lowest 3.6/4.5- and 3.6/5.8-\micron~ flux ratios in Figure \ref{fig_flux_ratios}.}
Furthermore, thermal H$_2$ emission with H+He collisions cannot account for the large [3.6]-[4.5] and [3.6]-[5.8] colors ($>$2.1 and $>$2.6, respectively) in IRAS 16547--4247, W 75 N, G 11.92--0.61 {and G 324.72+0.34}.
In Figure \ref{fig_cc_obs_vs_H2b} we compare the linear correlation of the observed [3.6]-[4.5] and [3.6]-[5.8] colors with modeled thermal H$_2$ emission for collisions with H$_2$+He (Figure  \ref{fig_cc_H2_models}) in more detail. The figure shows that the linear correlation observed in Green Fuzzy emission may be attributed to different densities of gas observed in thermal H$_2$ emission. The problems with this explanation are discussed in Section 5 in detail. 

\subsection{PAH and fluorescent H$_2$ emission}
Figure \ref{fig_cc_with_PDRs} shows color-color diagrams for the models of PAHs and fluorescent H$_2$ emission, usually associated with photodissociation regions \citep[PDRs; e.g.,][for a review]{Hollenbach97}. Their fluxes are obtained from \citet{DL07} and \citet{DB96}, respectively. 
The models by \citet{DL07} also include thermal dust continuum, which is prominently observed at longer wavelengths, but its contribution is negligible in the IRAC bands as compared with PAHs.
For comparison we selected their results for a UV field 10$^2$--10$^5$ as large as the interstellar radiation field for the solar neighborhood. Such a UV field is comparable to those observed in well-known dense PDRs associated with OB stars such as Orion Bar, NGC 2023 and 7023 \citep[e.g.,][]{Tielens08,Usuda96,Takami00}. For the abundance of PAH, all the results calculated for the dust/gas ratio in the Milky Way (fraction of C of 0.47--4.58 \% to the entire abundance in the interstellar medium) are included in the figure.

In Figure \ref{fig_cc_with_PDRs} the colors for PAHs are noticeably offset from the linear correlation of [3.6]-[4.5] and [3.6]-[5.8] colors, with a combination of small [3.6]-[4.5] colors and large [3.6]-[5.8]. This indicates that PAHs are not primarily responsible for the color distribution observed in Green Fuzzy emission. In contrast, the constant [3.4]-[5.8] colors observed in
%G 5.89--0.89 and
several positions in IRAS 20126+4104 are similar to those of PAHs. The figure shows that the modeled colors for fluorescent H$_2$ show similar correlation between the [3.6]-[4.5] and [3.6]-[5.8] colors to those observed in most of Green Fuzzies. The discrepancy between the modeled and observed colors could be attributed to an extinction $A_K$ up to $\sim$20. {This combination, however, cannot explain some pixels in G 35.2-0.7 N and G 324.72+0.34, with the lowest [3.6]-[4.5] and [3.6]-[5.8] colors.}
%and lower colors than the fluorescent H$_2$ models.}
We discuss the possible contribution of fluorescent H$_2$ emission to the observed colors in detail in Section 5.3.

\section{Discussion}

\subsection{Scattered continuum as a possible primary source of Green Fuzzy emission}

The colors of the star+disk flux via scattering and extinction are approximately consistent with the observed colors. In particular, a similar correlation between [3.6]-[4.5] and [3.6]-[5.8] colors are observed in scattered continuum in the cavity of the L 1527 protostellar outflow, agreeing with this explanation. In this case, the color distribution observed in each object can be attributed to differing foreground extinctions (including dense cores and envelopes). This is corroborated by the facts that (1) the position of the protostar matches the regions with the lowest $I_{3.6~\micron}$/$I_{4.5~\micron}$ and $I_{3.6~\micron}$/$I_{5.8~\micron}$ ratios for most of the objects (Figure \ref{fig_flux_ratios}); and (2) the column density of the gas and dust is higher close to the protostar than in surrounding regions (see references for Table \ref{target list}). This also suggests that the internal extinction in the extended emission regions is relatively small, otherwise the $I_{3.6~\micron}$/$I_{4.5~\micron}$  and $I_{3.6~\micron}$/$I_{5.8~\micron}$  ratios would decrease with increasing distance from the protostar due to reddening. In this context, the Green Fuzzy emission we analyzed may represent the morphology of cavities due to outflowing gas, and the $I_{3.6~\micron}$/$I_{4.5~\micron}$ and $I_{3.6~\micron}$/$I_{5.8~\micron}$  ratio maps in Figure \ref{fig_flux_ratios} may represent the distribution of foreground extinction.

The explanation that Green Fuzzy emission is scattered continuum associated with outflow cavities is consistent with the fact their distribution is more similar to the 
%$K$-band
{2-\micron}
continuum than to the H$_2$ 2.12 \micron~ emission (Section 2). The morphological discrepancies between the 4.5-\micron~{emission and continuum at 2 \micron} could be attributed to different extinctions at these wavelengths. This explanation is also consistent with the fact that such emission in some objects are associated with signatures of molecular outflows and/or outflow shocks \citep[Section 2; see also][]{Shepherd07,Araya07,Qiu08,Cyganowski09,Cyganowski11}. It is intriguing that the distribution of the IRAC emission and/or 
%$K$-band 
{2-\micron}
continuum do not match H$_2$ emission and/or molecular outflows in some objects, including IRAS 05358+3543, IRAS 16547--4247 and W 75 N (Section 2). This would indicate the complicated nature of high-mass star forming regions, which are often associated with multiple protostars and outflows like these objects \citep[e.g.,][]{Shepherd00,Shepherd03,Beuther02,Franco09}.

As shown in Sections 3 and 4.2, some positions in G 192.16--3.82, G 35.2--0.7 N, and W 75 N show a [3.5]-[5.8] excess from the linear correlation of the [3.6]-[4.5] and [3.6]-[5.8] colors in the remaining region. This may be attributed to a different origin than scattered continuum. The deviation of the color observed in G 192.16--3.82 approximately matches the colors of PAHs in Figure \ref{fig_cc_with_PDRs}, suggesting that this is due to diffuse contaminating PAH emission in the outer region where Green Fuzzy emission is relatively faint (Figure \ref{fig_deviation1}). This explanation is corroborated by the fact that the objects near G 192.16--3.82 are associated with extended emission at 8.0-\micron~(Figures 1 and 2), a signpost of PAH emission (Section 1). The same explanation may also apply to the [3.6]-[5.8] excess in W 75 N. This is corroborated by the fact that the region associated with the [3.6]-[5.8] excess is close to extended 8.0-\micron~emission (Figures 1 and 2) associated with an H II region \citep{Haschick81}.

An alternative possible contributor for W 75 N and G 35.2--0.7 N is shocks the physical conditions of which are similar to some of those associated with low-mass protostars. This is consistent with the fact that the [3.6]-[5.8] excess in these objects is observed over color ranges similar to such shocks in Figure \ref{fig_cc_HH}. Note that the fraction of the 4.5-\micron~flux in this component to that of the entire region (22, 34, 9 \% for G 192.16--3.82, G 35.2--0.7 N, and W 75 N; Sections 3 and 4.2) is an upper limit for the additional emission component, assuming that the flux in the marked regions is solely due to shocks or PAHs. Throughout, the contaminating emission is a minor component in each object in terms of both spatial coverage and flux. 

\subsection{Thermal H$_2$ emission as a possible primary source of Green Fuzzy emission}
As described in Section 1, H$_2$ or CO emission in shocks is often regarded as the primary source of Green Fuzzy emission \citep{Rathborne05, Araya07, Shepherd07, Qiu08, Morales09,Cyganowski08,Cyganowski09, Cyganowski11,Chambers09}. Such emission is usually expected via thermal excitation. Indeed, thermal H$_2$ and CO emission explain well the IRAC colors observed in shocks associated with low-mass protostars \citep{Ybarra09,Takami10b}. However, the [3.6]-[4.5] and [3.6]-[5.8]  colors show different correlations between Green Fuzzy emission and these shocks associated with low-mass protostars (Section 4.1, Figure  \ref{fig_cc_HH}). This implies that different shock conditions are required if Green Fuzzy emission is due to shocks.

As described in Section 4, the Green Fuzzy emission in our sample does not show any clear signature for the presence of the ro-vibrational CO emission, which should only appear in the 4.5-\micron~band. In this subsection we therefore focus our discussion on thermal H$_2$ emission, which is primarily responsible for shocked H$_2$ emission in most cases  \citep[e.g.,][]{Brand88, Gredel94, Everett95, Richter95,Usuda96, Eisloffel00,Takami06a,Beck08,Neufeld06,Neufeld08,Neufeld09}. Indeed, the colors observed at most of the positions in Green Fuzzy emission are consistent with thermal H$_2$ emission (Section 4.1, Figure \ref{fig_cc_obs_vs_H2}), which is primarily responsible for shocked H$_2$ emission. Furthermore, the linear correlation of the [3.6]-[4.5] and [3.6]-[5.8] colors may be attributed to different densities if thermal H$_2$ emission with H$_2$ and He collisions is responsible for the Green Fuzzy emission (Section 3.1, Figure \ref{fig_cc_obs_vs_H2b}).

These explanations, however, face several problems described below.
First,  in {six out of nine objects} ({IRAS 05358+3543}, IRAS 16547--4247, G 35.2--0.7 N, G192.16--3.82,W 75 N, G298.26-0.74), some positions show smaller [3.6]-[4.5] and [3.6]-[5.8] colors ($<$1.0 and $<$1.5, respectively) than those predicted by thermal H$_2$ emission (Section 4.1, Figure \ref{fig_cc_obs_vs_H2}). In these objects the linear correlation between [3.6]-[4.5] and [3.6]-[5.8] is observed across both color ranges, suggestive of another origin.
Secondly, Figures \ref{fig_cc_obs_vs_H2} and \ref{fig_cc_obs_vs_H2b} would indicate the temperature of thermal H$_2$ is $\sim$2000 K, in which H$_2$ should be efficiently excited for the  1-0 S(1) 2.12 \micron~emission \citep[e.g.,][]{Brand88, Gredel94, Everett95, Richter95,Usuda96, Eisloffel00,Takami06a,Beck08,Neufeld06,Neufeld08,Neufeld09}. It would therefore be puzzling if the distribution of Green Fuzzy emission were remarkably different from H$_2$ 2.12 \micron~emission (Section 2). {See also \citet{Takami10b} for the dependence on density and temperature of H$_2$ emission at the 2.12 \micron~and IRAC bands.}

Thirdly, if this color correlation is due to the different densities in Figure \ref{fig_cc_obs_vs_H2b}, it implies that the positions with larger [3.6]-[4.5] and [3.6]-[5.8] colors (thereby lower $I_{3.6\micron}/I_{4.5\micron}$ and $I_{3.6\micron}/I_{5.8\micron}$ flux ratios) are associated with lower densities. Since the $I_{3.6\micron}/I_{4.5\micron}$ and $I_{3.6\micron}/I_{5.8\micron}$ flux ratios increases with distance in G 35.2--0.7 N, G 192.16--3.82, W 75 N, 
the density should increase with distance in these objects. 
This is in the opposite sense as one would expect in outflow shocks. If these shocks are due to interaction between the ejecta and surrounding gas, as in many case \citep[e.g.,][]{Bally07,Arce07}, one would expect lower densities in the downstream as for the distribution of molecular gas around protostars \citep[e.g.,][]{Stahler05}. If these shocks are in the ejecta, one would also expect such a density distribution as long as the flow has an opening angle similar to that observed in many molecular outflows \citep[e.g.,][]{Beuther05,Arce07}.
%Furthermore, the IRAC flux for thermal H$_2$ emission should increase with density at the temperatures we discuss \citep{Takami10b}, hence one may expect the lowest flux closest to the protostar in these regions. This is, however, the opposite trend from the observed flux distribution (Section 3, Figure \ref{fig_3color_1D})

Fourthly, it is likely that the ro-vibrational CO emission significantly contributes to the entire 4.5-\micron~flux if the hydrogen number density exceeds $\sim10^7$ cm$^{-3}$ \citep[][]{Neufeld08, Takami10b}. This would cause larger [3.6]-[4.5] colors at high densities, as observed in colors in shocks associated with low-mass protostars at [3.6]-[5.8] $\sim$1.5-2.0 \citep[Figure \ref{fig_cc_HH}; see][for details]{Takami10b}. However, such a tendency is not seen for the colors of Green Fuzzy emission in Figure \ref{fig_cc_obs_vs_H2b}.

Throughout, we conclude that it is not likely that Green Fuzzy emission is primarily due to thermal H$_2$ emission. {One might think that a combination of thermal H$_2$ emission and scattered continuum could explain the linear correlation we have discussed. This explanation could overcome the first problem described above, interpreting the lowest [3.6]-[4.5] and [3.6]-[5.8] colors as scattered continuum. 
However, the second problem still remains. The third and fourth problems could only be overcome if the shock conditions were uniform, yielding a single combination of [3.6]-[4.5] and [3.6]-[5.8] colors over the region. This is not likely, considering the fact that shocks associated with high-mass protostellar outflows show complicated shock structures \citep[see, e.g., ][]{Kaifu00,Davis07,Cunningham09}, suggesting the presence of different physical conditions in each outflow.
It is noteworthy that, according to this explanation, we would expect scattered continuum further away from the protostar than H$_2$ emission (i.e., the regions with the largest $I_{3.6\micron}/I_{4.5\micron}$ and $I_{3.6\micron}/I_{5.8\micron}$) in G 35.2-0.7 N. 
This is the opposite trend as observed in the H$_2$ 2.12-\micron~emission and 
scattered continuum at 2 \micron~\citep{Froebrich11,Lee12}. There is no clear theory for how the above explanation overcomes this discrepancy.

Alternatively, a combination of thermal and fluorescent H$_2$ emission in shocks may explain the linear correlation in the color-color diagrams. We discuss this in the next subsection.}

\subsection{Fluorescent H$_2$}

The linear correlation of the [3.6]-[4.5] and [3.6]-[5.8] colors may also be explained by a combination of fluorescent H$_2$ emission and extinction (Section 4.3). To investigate this possibility in detail, we calculate the IRAC fluxes expected for fluorescent H$_2$  in PDRs based on \citet{DB96}. As for Section 4.3 and Figure  \ref{fig_cc_with_PDRs}, we selected their results for a UV field 10$^2$--10$^5$ as large as the interstellar radiation field for the solar neighborhood. Again, such a UV field is comparable to those observed in well-known dense PDRs associated with OB stars such as Orion Bar, NGC 2023 and 7023 \citep[e.g.,][]{Tielens08,Usuda96,Takami00}.
%Among their models, we select all the results for the UV field relative to the averaged solar neiborhood ($\chi$) of 10$^2$ or larger, up to 10$^5$. This range is comparable to those for dense PDRs associated with young OB stars \citep[e.g.,][]{Tielens08, Usuda96, Takami00}. 
These are tabulated in Table \ref{tbl-fluH2-PAH}. The table shows that the predicted fluxes are up to $\sim$3 MJy str$^{-1}$, significantly lower than those observed in the bright areas of the Green Fuzzy emission shown in Figure \ref{fig_3color_1D} ($\gg$ 10 MJy str$^{-1}$).

Furthermore, PDRs are also associated with the PAH emission, which appears to be much brighter in the 3--8 \micron~range \citep{Tielens08}. To investigate this issue more quantitatively, we calculate the modeled fluxes for the same range of UV fields based on \citet{DL07}. \citet{DL07} provide fluxes per column density, without the models for the PDRs. We therefore assume a column density for the PDRs of A$_V$=1, a typical column density for PDRs \citep[e.g.,][]{Hollenbach97}, corresponding to $1.9 \times 10^{21}$ cm$^{-2}$ \citep[][]{Mathis00}. These are also tabulated in Table \ref{tbl-fluH2-PAH}, for minimum and maximum abundances of PAHs modeled by \citet{DL07} ($q_{PAH}$=0.47 and 4.58 \%, respectively). The larger $q_{PAH}$ value is comparable to those of well-studied dense PDRs \citep[$\sim$3.5 \%,][]{Tielens08}. Table \ref{tbl-fluH2-PAH} shows that, even for the minimum abundance of PAH (i.e., lower than those of well studied dense PDRs by a factor of $\sim$7), the modeled PAH fluxes are larger than fluorescent H$_2$ by a factor of $>$6 and $>$100 at 3.6 and 5.8 \micron, respectively. Thus, despite the uncertainty in the calculations (i.e., the column density of the PDRs), it is not likely that the fluorescent H$_2$ emission dominates over the PAH emissions in normal PDRs.

Fluorescent H$_2$ emission has been observed in the high-mass protostellar outflow DR 21 \citep{Fernandes97, SmithH06} and a well-known Herbig-Haro object \citep{Fernandes95}. A combination of thermal and fluorescent H$_2$ may consistently explain the linear correlation of [3.6]-[4.5] and [3.6]-[5.8] colors observed in Green Fuzzy emission in most of the objects (Figures \ref{fig_cc_obs_vs_H2}, \ref{fig_cc_obs_vs_H2b}, \ref{fig_cc_with_PDRs}), if most of the PAHs+dust grains are destroyed by the passage of shocks. \citet{Fernandes97} and \citet{SmithH06} suggested a UV field of H$_2$ excitation of $\chi \sim 10^2 - 10^4$ in the DR 21 outflow. Again, such a UV field would yield H$_2$ flux too low compared to the observations. This problem may be  overcome if (1) most of the dust grains are destroyed by shocks, allowing UV photons to be absorbed more efficiently by molecular hydrogen; and/or (2) turbulence in the shocks make the ultraviolet H$_2$ lines broader, allowing more photons to be absorbed. Models for the coexistence of fluorescent and thermal H$_2$ in shocks are required to further investigate this possibility. Alternatively, follow-up spectroscopic observations of Green Fuzzy emission would allow us to further constrain the associated shock conditions.

%While infrared H$_2$ emission in shocks are thermal in many cases\citep[e.g.,][]{Brand88, Everett95, Richter95,Usuda96, Eisloffel00,Takami06a,Beck08}, 

\subsection{PAHs}
Our color-color diagrams show that PAHs are not the primary emission mechanism for Green Fuzzy emission (Section 4.3, Figure \ref{fig_cc_with_PDRs}). %In this section we discuss the cases for G 5.89--0.39 and IRAS 20126+4104, in which a relatively constant [3.6]-[5.8] color is observed in the entire region and several positions, respectively.
In this section we discuss the cases for IRAS 20126+4104, in which a relatively constant [3.6]-[5.8] color is observed at several positions.

%The entire region of G 5.89--0.39 and several positions in IRAS 20126+4104 show a relatively constant [3.6]-[5.8] color
These positions in IRAS 20126+4104 show a relatively constant [3.6]-[5.8] color
%(2.6--3.3 and $\sim$2.4, respectively)
similar to PAHs (Section 4.3, Figure \ref{fig_cc_with_PDRs}).
%Indeed, PAHs with a variety of conditions show a relatively narrow range of [3.6]-[5.8] colors in Figure \ref{fig_cc_with_PDRs}.
While we do not reject the possibility that these emission components are primarily due to thermal H$_2$ (Section 4.1, Figures \ref{fig_cc_obs_vs_H2}, \ref{fig_cc_obs_vs_H2b}), this coincidence suggests that at least their 3.6- and 5.8 \micron~ fluxes are primarily due to PAHs. However, the observed [3.6]-[4.5] colors in these regions are significantly larger than those of PAHs modeled by \citet{DL07}, 
%ranging from 0.3 to 2.0.
ranging from 0.8 to 1.5. This may be primarily due to a contribution from another emission mechanism at 4.5-\micron, in which the PAHs emission is the faintest among the four IRAC bands \citep[e.g.,][see also Table \ref{tbl-fluH2-PAH}]{Reach06,DL07,Tielens08}.

In IRAS 20126+4104, this emission component is associated with a point source to the west of the high-mass protostar (Section 3, Figure \ref{fig_deviation1}). The three-color image in Figure  \ref{fig_IRAC_vs_obsH2} shows the presence of excess emission at 8.0-\micron, supporting the idea that this region is associated with PAH emission (Section 1). This region is surrounded by emission with different colors, which can be attributed to either scattered continuum (Section 5.1) or H$_2$ emission (Sections 5.2, 5.3). Thus, the different [3.6]-[4.5] colors described above can be naturally attributed to different contributions from two emission components (i.e., PAH and scattered continuum, or PAH and H$_2$ emission).
%In G 5.89--0.39, the south-east part of the emission region shows large $I_{3.6\micron}/I_{4.5\micron}$ (Figure \ref{fig_flux_ratios}), and therefore small [3.6]-[4.5] values close to the colors of modeled PAHs (Figure \ref{fig_cc_with_PDRs}). This part of the region also appears red in the three-color images in Figure \ref{fig_3color_1D}, indicative of the presence for excess emission at 8.0 \micron. These suggest that
%the south-east part of the emission is mainly due to PAHs. In contrast, the northwest region (i.e., the region close to the high-mass protostar) has significantly larger [3.6]-[4.5] colors than PAHs, indicative of contribution from another emission mechanism, in particular at 4.5-\micron. Again, this may be due to either scattered continuum (Section 5.2) or H$_2$ emission (Sections 5.1 and 5.3).

%Several of
Four of
our targets are associated with hypercompact and ultracompact H II regions (Table \ref{target list}). Thus, these should also be associated with far-UV radiation which can excite PAHs. However, none of our targets show clear evidence for the presence of PAH emission directly associated with the protostar. These suggest that the high-mass protostars and hypercompact/ultracompact H II regions are heavily embedded even at the wavelengths of our analysis, i.e., 3.6--5.8 \micron.
%In the case of G 5.89--0.39, the diameter of the ultracompact H II region is $\sim$5'' \citep{Puga06,Hunter08}, nearly the same as the saturated region where we have not been able to apply analysis with flux ratios and color-color diagrams (Figures \ref{fig_3color_1D}, \ref{fig_flux_ratios}).

\section{Caveats}
It is beyond the scope of this paper to discuss the emission mechanisms below, but we briefly state their possible contributions.

Thermal dust continuum might also contribute to the 3.6-5.8 \micron~flux. Such emission can be extended in the outflow cavity \citep[e.g.,][]{DeBuizer06}. According to \citet{DL07} this emission component is negligible compared with PAH emission in the IRAC bands. However, the contribution from the thermal dust continuum may not be negligible in circumstances where PAHs are destroyed but larger grains survive.

Fast shocks should be associated with atomic and ionic lines, such as Pf $\gamma$ 3.74 \micron, Br $\alpha$ 4.05 \micron, Pf $\beta$ 4.65 \micron, and Fe II 5.34 \micron~\citep[][{see Table \ref{line list} for details}]{Reach06}, and these may also contribute to the 3.6-5.8 \micron~bands.
%Such shocks would produce UV radiation which may cause fluorescent H$_2$.
In the limited cases discussed by \citet{Reach06} the above lines in shocks yield
{$I_{3.6\micron}/I_{4.5\micron}$} and $I_{3.6\micron}/I_{5.8\micron}$
ratios of $\la 0.1$, significantly lower than those observed in Green Fuzzies (Figure \ref{fig_flux_ratios}). Thus, this possibility should be addressed with more detailed modeling and/or spectroscopic observations.

\section{Conclusions}

We analyzed the flux ratios and pixel-pixel colors of  ``Green Fuzzy" emission
toward {six} nearby  {($d$=2--3 kpc)} well-studied high-mass protostars and three candidate
high-mass protorstars observed using the archival data for Spitzer IRAC at 3.6, 4.5, and 5.8 \micron.
In color-color diagrams most of the sources show a positive correlation
between the [3.6]-[4.5] and [3.5]-[5.8] colors, in all or a part of the region,
along the extinction vector. We compare them with the colors of other objects and models, i.e.,
(1) modeled scattered continuum, and observed scattered continuum associated with the L 1527 low-mass
      protostellar outflow;
(2) shocks associated with low-mass protostars and modeled thermal H$_2$ emission;
(3) modeled emission for fluorescent H$_2$ ; and
(4) modeled emission for PAHs.
The conclusions are as follows:-

\begin{enumerate}
\item Of the above emission mechanisms, scattered continuum in the outflow cavities provides the simplest explanation
for the observed linear correlation. Indeed, such a correlation is also observed in the scattered continuum in L 1527. In this case, the different
colors within each object are attributed to different degrees of foreground extinction at different positions in the extended emission. Different objects show different colors not only along the extinction vector, but also across it. This can be attributed to different intrinsic colors of the star(+disk) system between the objects.
This interpretation, that scattered continuum is responsible for Green Fuzzy emission, is consistent with the fact that the distribution of emission is remarkably different from H$_2$ 2.12 \micron~emission in some objects, but more similar to the 
%$K$-band
continuum at 2 \micron.

\item The observed color correlation is remarkably different from that observed in molecular shocks in low-mass protostars, in which the emission is due to thermal H$_2$, plus CO at high densities. Even so, a significant fraction of the Green Fuzzy emission show colors within the color range of modeled thermal H$_2$ emission. However, this emission mechanism is not likely for the following reasons: (1) the observed linear correlation in some objects exceeds the color range predicted for thermal H$_2$; (2) while ro-vibrational CO emission should contribute to the 4.5-\micron~emission at high densities ($n_H \ga 10^7$ cm $^{-3}$), the observed color-color diagrams do not clearly show such a signature; (3) the temperature inferred by the color diagrams ($\sim$2000 K) would allow for H$_2$ 2.12\micron~emission to be observed, however, the distribution of this line emission does not match Green Fuzzy emission in some objects; and (4) higher densities in the outer region compared to the inner region are required to explain the color correlation in some objects, however, this is not likely.\\
The first problem may be overcome if we attribute the lowest [3.6]-[4.5] and [3.6]-[5.8] colors to another origin, e.g., scattered continuum. However, a combination of thermal H$_2$ emission and scattered continuum does not seem to overcome the other problems described above.

\item Fluorescent H$_2$ in normal PDRs cannot account for the observed flux. Furthermore, emission from PAHs should dominate over fluorescent H$_2$ in the IRAC bands in such circumstances. Fluorescent H$_2$ in shocks may overcome these problems if shocks destroy PAHs and dust grains, and turbulence in shocks allows far-UV photons to be absorbed by H$_2$ more efficiently. A combination of fluorescent and thermal H$_2$ may also explain the observed linear correlation between [3.6]-[4.5] and [3.6]-[5.8] colors. Shock models with these excitation mechanisms are required to investigate the feasibility of this explanation.

\item As expected, our color-color diagrams show that PAHs are not the primary mechanism for Green Fuzzy emission. PAHs may significantly contribute in
%G 5.89--0.39, and
a part of the emission regions in another few objects including IRAS 20126+4104, G 192.16--3.82 and W 75 N. None of our sample show clear evidence for PAH emission directly associated with the high-mass protostars, several of which should be associated with ionizing
radiation. This suggests that those protostars are heavily embedded even at mid-infrared.
\end{enumerate}

In summary, among the emission mechanisms discussed above, scattered continuum in outflow cavities provides the simplest explanation
for the observed linear correlation. Alternative possible emission mechanisms to explain the above linear correlation may be a combination of thermal and fluorescent H$_2$ emission in shocks, {and a combination of scattered continuum and thermal H$_2$ emission,} but detailed models or spectroscopic follow-up are required to further investigate this possibility. The {first} interpretation {(i.e., attributing the Green Fuzzy emission to the scattered continuum in outflow cavities)} does not change interpretations of some previous literature which show that Green Fuzzy emission is associated with signatures of jets and outflows (e.g., molecular outflows, shock tracer, or collimated ionized jet).

This work also highlights the difficulty in obtain definitive conclusions for the excitation mechanism of Green Fuzzies. Spectroscopic follow-up would allow us to complement this study in order to understand the nature of high-mass protostars themselves, or conditions of outflow shocks associated with high-mass protostars in detail.

\acknowledgments
We first thank two anonymous referees for careful reviews and useful discussions.
We are grateful for Drs. Ho, Hirashita and Hirano for useful discussions. The IRAC images were obtained through the Spitzer archive operated by IPAC.
This research made use of the Simbad data base operated at CDS, Strasbourg, France, and the NASA's Astrophysics Data System Abstract Service. This study is supported from National Science Council of Taiwan (Grant No. {98-2112-M-001 -002 -MY3 and 100-2112-M-001-007-MY3}).


%% in parentheses, after the keyword, but they will not be verified.

{\it Facilities:} \facility{Spitzer Space Telescope (IRAC)}.

\clearpage

%% the file, uncomment the \plotone, \plottwo, and
%% If you need a layout that cannot be achieved with \plotone or
%% \plottwo, you can invoke the graphicx package directly with the
%% \includegraphics command or use \plotfiddle. For more information,
%% please see the tutorial on "Using Electronic Art with AASTeX" in the
%% documentation section at the AASTeX Web site,
%% http://www.journals.uchicago.edu/AAS/AASTeX.
%%
%% The examples below also include sample markup for submission of
%% supplemental electronic materials. As always, be sure to check
%% the instructions to authors for the journal you are submitting to
%% for specific submissions guidelines as they vary from

%% This example uses \plotone to include an EPS file scaled to
%% 80% of its natural size with \epsscale. Its caption
%% has been written to indicate that additional figure parts will be
%% available in the electronic journal.

%%% Fig. 1: three-color images and x-y cutouts %%%
% I05	97%
% I16	96 %
% G35	98 %
% G192	98 %
% I20	97 %
% G5	97 % 
% W75	99.5 %
% G11	98 %
% G 298	98 %
% G 324   98 % < revised
% L1527  98 % < revised
%

\clearpage

%%% Fig. 2: comparison with H2 2.12 micron in I05, I20 and G 5.89 %%%

\clearpage

%% Here we use \plottwo to present two versions of the same figure,
%% one in black and white for print the other in RGB color
%% for online presentation. Note that the caption indicates
%% that a color version of the figure will be available online.
%%

%%% Fig. 3: background subtraction %%%

%%% Fig. 4: flux ratio maps %%%

\clearpage

%%% Fig. 5: color-color diagrams (objects only) %%%

%%% Fig. 6: deviation of colors in G 192.16--3.82, IRAS 20126+4104 %%%
\begin{figure*}
\epsscale{2.2}
%\epsscale{1.0}
\plotone{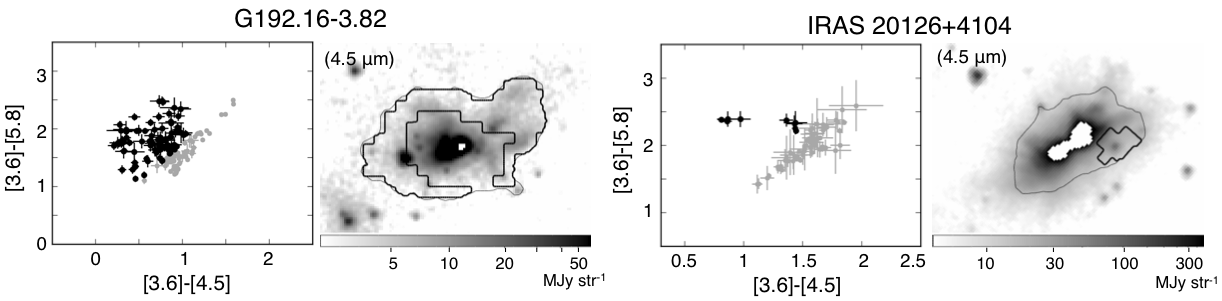}
%\plotone{f6.eps}
\caption{
Emission deviating from the linear color correlation in the G 192.16--3.82 and IRAS 20126+4104 regions, and their spatial distributions in the 4.5-\micron~image. The black and gray dots in the color-color diagrams show the deviating and remaining components of the [3.6]-[4.5] and [3.6]-[5.8] colors. In the 4.5-\micron~image the former is shown with thick curves, while the entire region used for the color-color diagrams are shown with thin curves. 
\label{fig_deviation1}}
\end{figure*}

%%% Fig. 7 (old Fig. 11): color-color diagrams (with L 1527) %%%
\begin{figure*}
\epsscale{2.2}
%\epsscale{1.0}
\vspace{-2cm}
\plotone{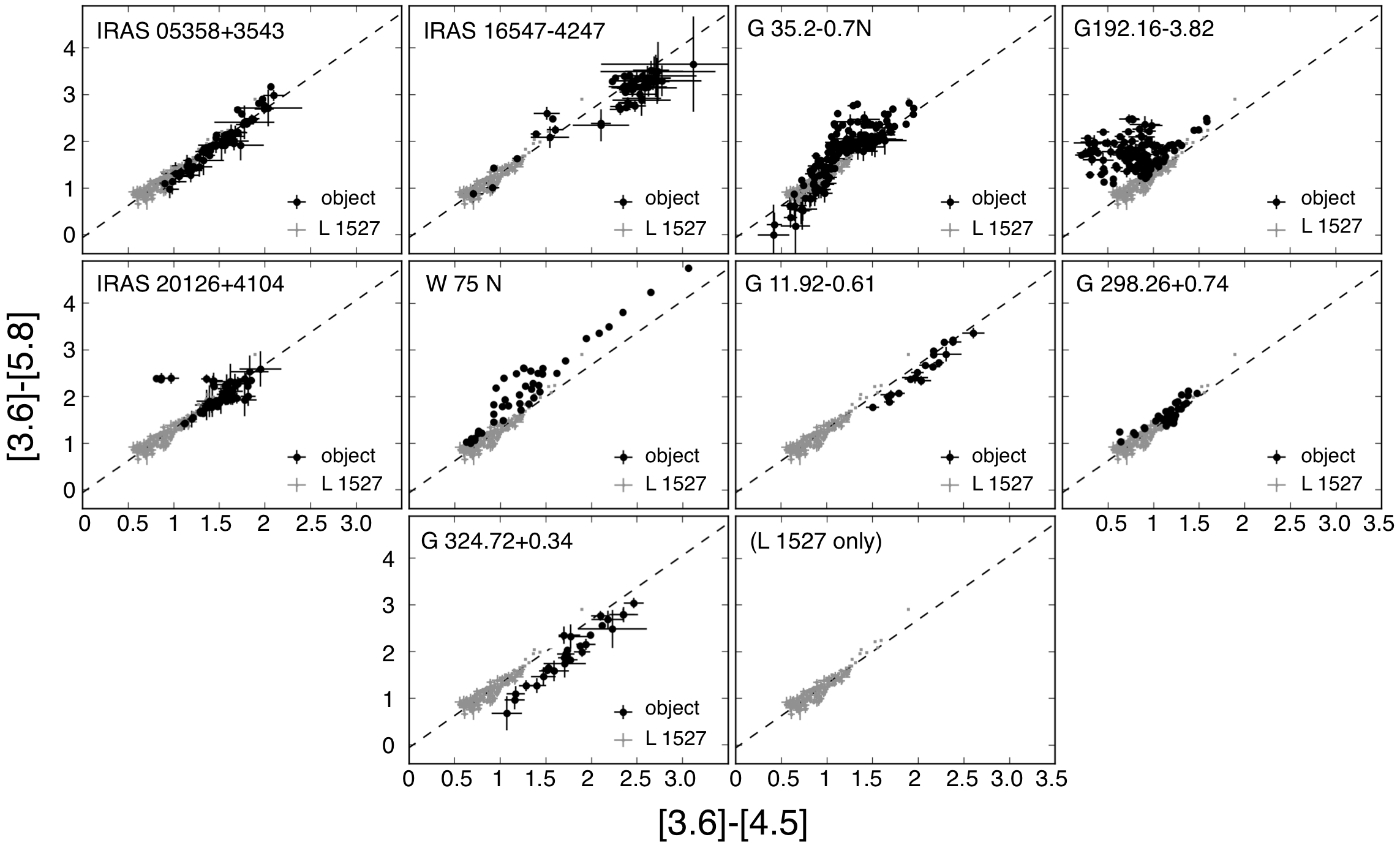}
%\plotone{f7.eps}
%\vspace{-1.5cm}
\caption{
Same as Figure \ref{fig_cc_hmonly} but with colors measured in the scattered continuum in the L 1527 outflow. The dashed line shows the regression line for the results of L 1527 ([3.6]-[5.8]=[3.6]-[4.5]$\times$1.374--0.067).
\label{fig_cc_with_L1527}}
\end{figure*}

%%% Fig. 8 (old Fig. 12): color-color diagrams (with star+disk models) %%%
\begin{figure*}
\epsscale{2.2}
%\epsscale{1.0}
\vspace{-2.5cm}
\plotone{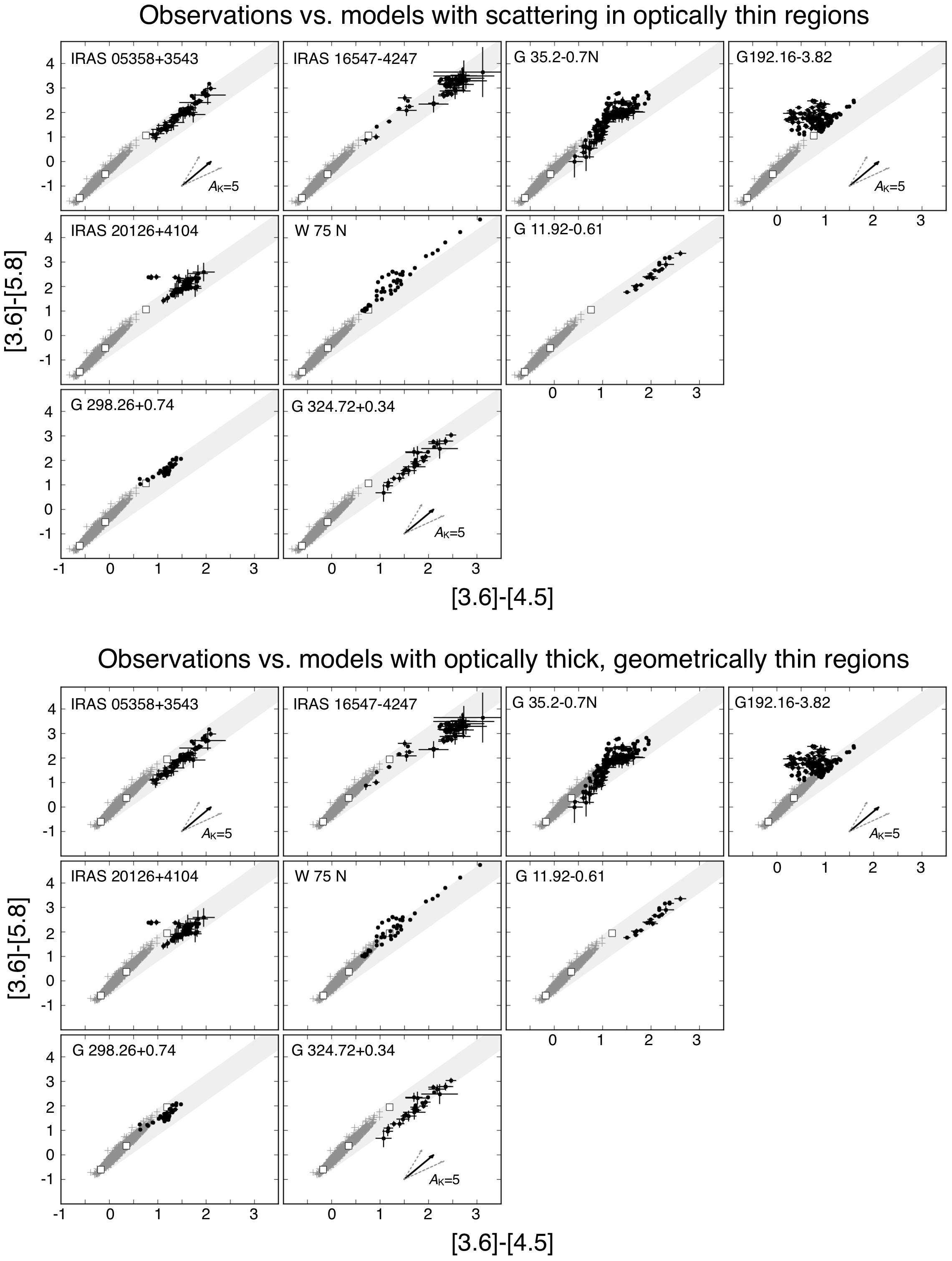}
%\plotone{f8.eps}
\caption{
Same as Figure \ref{fig_cc_hmonly} but with scattering for models of the optically thin cases (upper) and optically thick but geometrically thin cases (lower). The colors for the objects are shown with black dots,  with error bars if they are larger than the dots. Gray crosses show the colors for star-disk systems (Robitallile et al. 2006) with scattering. Open squares show the colors for a blackbody ($T$=4000, 1000 and 500 K from the lower-left to the upper-right) with scattering. The gray areas show the modeled colors, adding arbitrary extinction, assuming that the linear correlation observed in L 1527 is due to different extinction between positions.
\label{fig_cc_with_scamodels}}

\end{figure*}

\clearpage

%%% Fig. 9 (old Fig. 13): deviation of colors in G 192.16--3.82, IRAS 20126+4104 %%%
\begin{figure*}
\epsscale{2.2}
%\epsscale{1.0}
\plotone{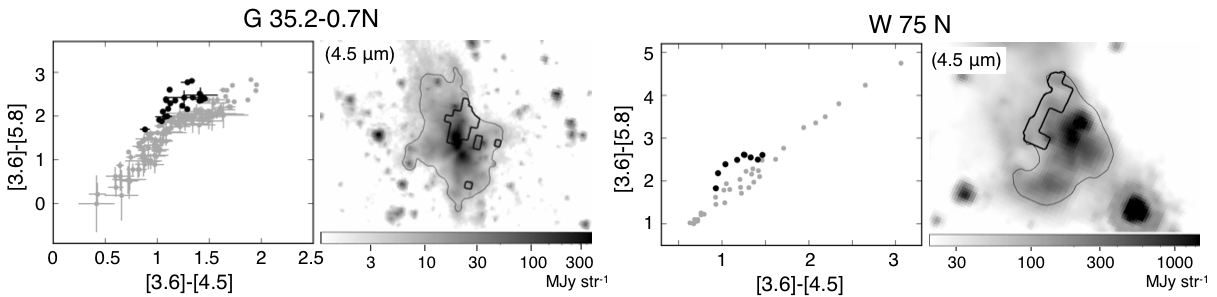}
%\plotone{f9.eps}
\caption{
Same as Figure \ref{fig_deviation1} but for G 35.2--0.7 N and W 75 N regions.
\label{fig_deviation2}}
\end{figure*}

%%% Fig. 10 (old Fig. 7): color-color diagrams (with low-mass protostellar jets) %%%
\begin{figure*}
\epsscale{2.2}
%\epsscale{1.0}
\plotone{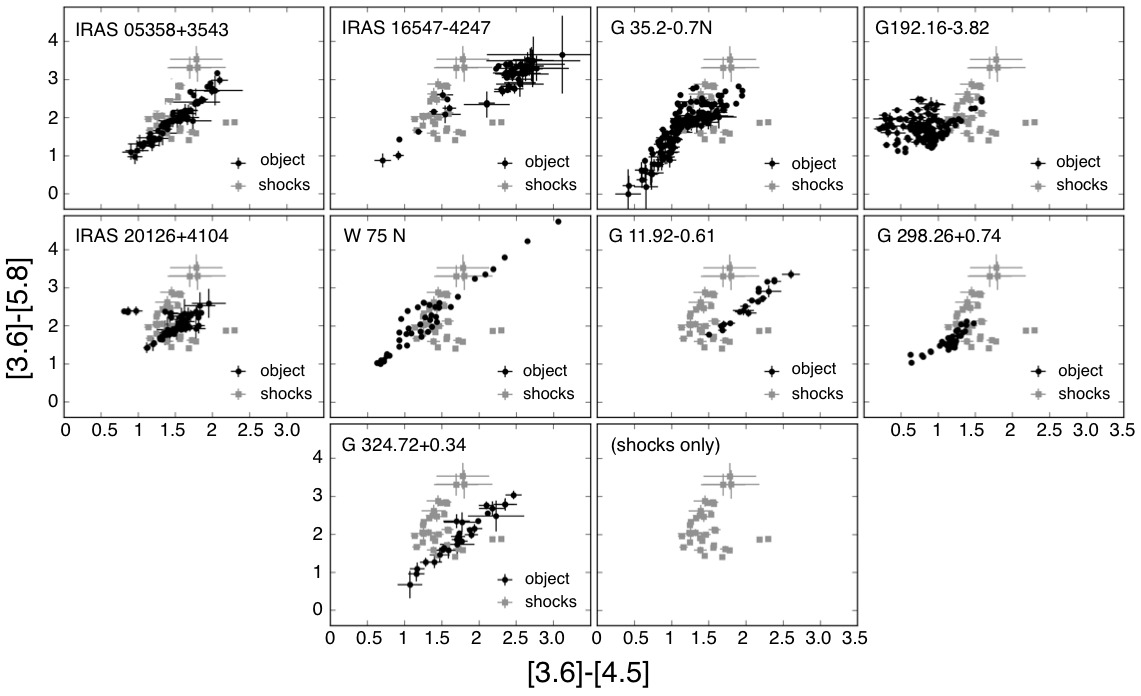}
%\plotone{f10.eps}
\caption{
Same as Figure \ref{fig_cc_hmonly} but with those observed in shocks associated with low-mass protostars. The error bars are shown only if they are larger than the dots.
\label{fig_cc_HH}}
\end{figure*}

%%% Fig. 11 (old Fig. 8): color-color diagrams (models for  thermal H2 emission) %%%
\begin{figure*}
\epsscale{2.2}
%\epsscale{1.0}
\plotone{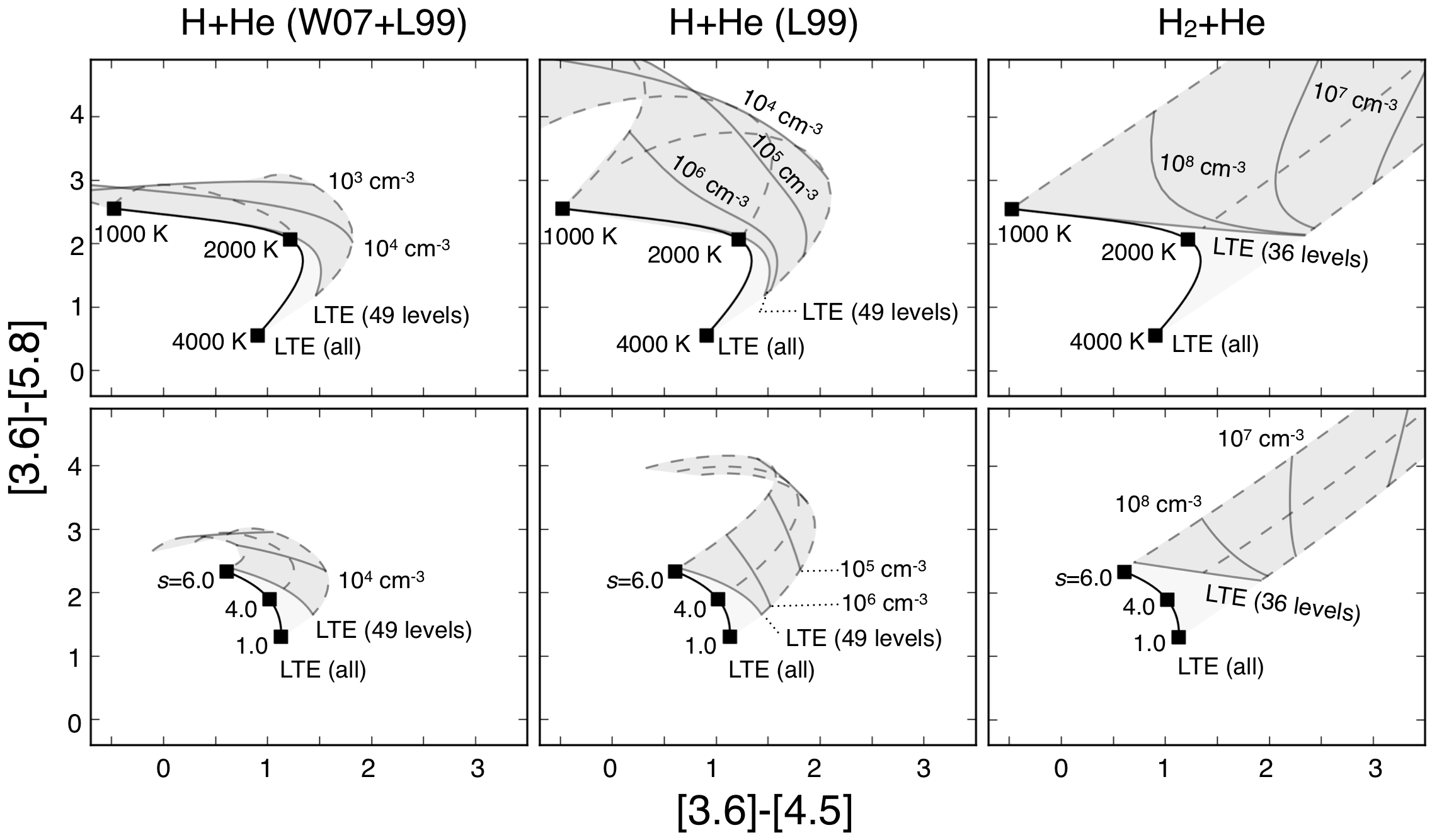}
%\plotone{f11.eps}
\caption{
Colors of thermal H$_2$ emission for LTE with all the transitions (solid curves) and non-LTE with 36/49 levels (dark gray area). The upper and lower figures are for isothermal cases and temperature structures determined by a power-law cooling function ($\Lambda \propto T^s$), respectively. The left plots are for collisions with H+He based on the collisional rate coefficients of H and He by Wrathmall et al. (2007) and Le Boutlot et al. (1999), respectively; middle are the same but the coefficients of H and He both by Le Boutlot et al. (1999); right are the same as middle but for collisions with H$_2$ and He. Gray solid and dashed curves show colors with the same density and temperature (or temperature structure determined by the power-law index), respectively. The light gray areas show the gap between LTE calculations for all and those with limited numbers of the transitions (see text).
\label{fig_cc_H2_models}}
\end{figure*}

%%% Fig. 12 (old Fig. 9): color-color diagrams (with thermal H2 emission) %%%
\begin{figure*}
\epsscale{2.2}
%\epsscale{1.0}
\vspace{-1.5cm}
\plotone{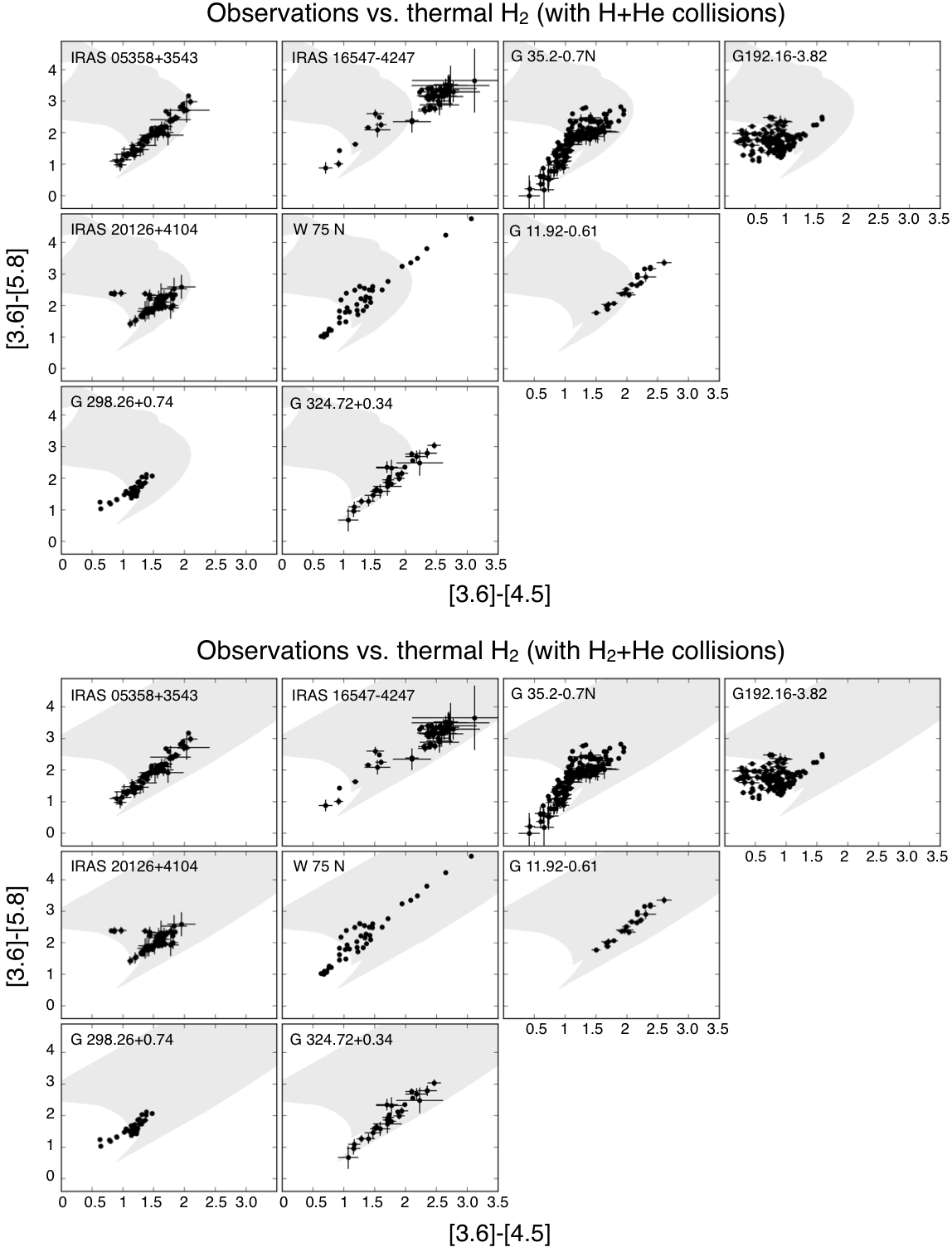}
%\plotone{f12.eps}
\caption{
Same as Figure \ref{fig_cc_hmonly} but with the possible coverage for thermal H$_2$ emission (gray area) shown in Figure \ref{fig_cc_H2_models}.
\label{fig_cc_obs_vs_H2}}
\end{figure*}

%%% Fig. 13 (old Fig. 10): color-color diagrams (with thermal H2 emission) %%%
\begin{figure*}
\epsscale{2.2}
%\epsscale{1.0}
\plotone{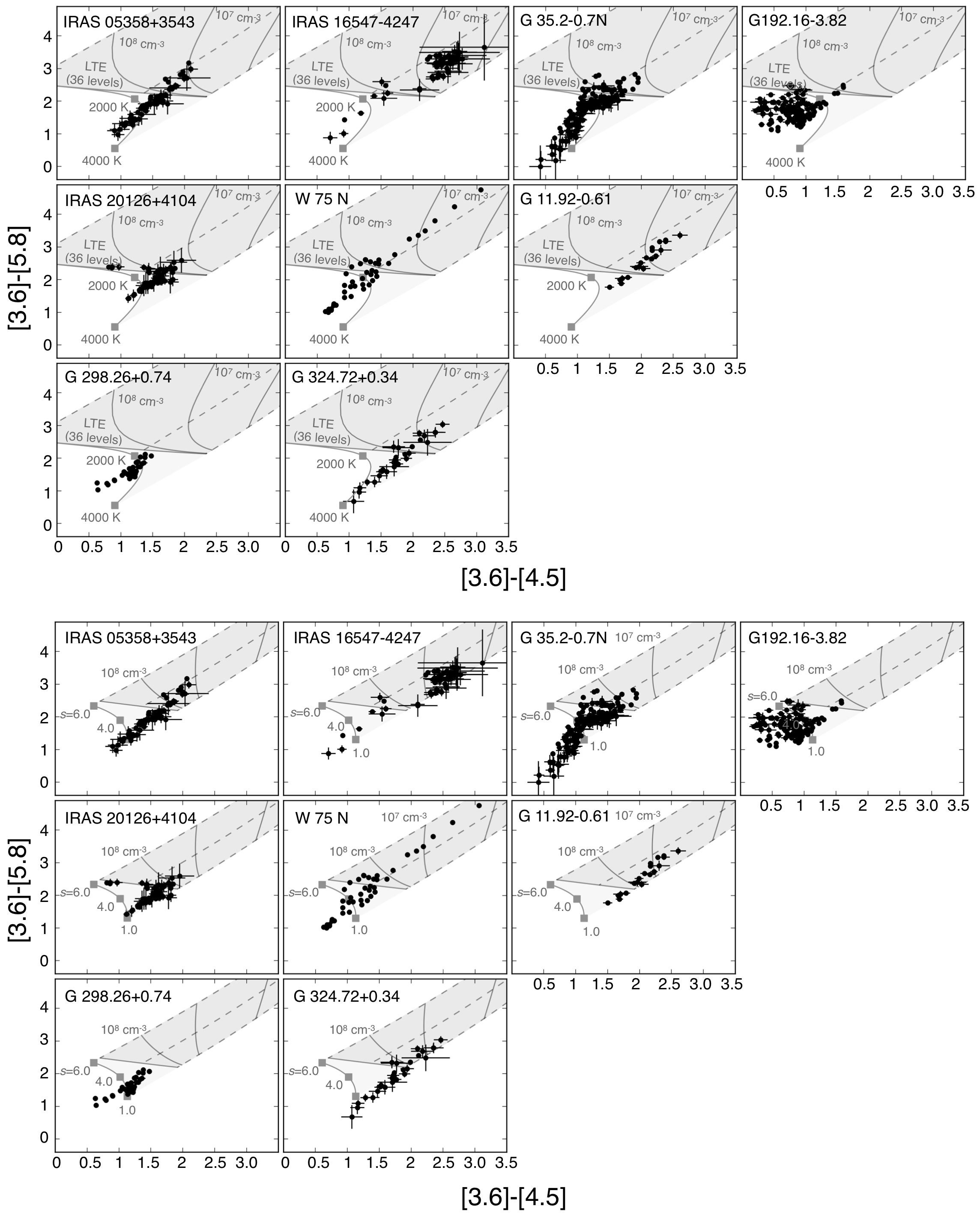}
%\plotone{f13.eps}
\caption{
Same as Figure \ref{fig_cc_hmonly} but with diagrams for modeled thermal H$_2$ emission for collisions with H$_2$ and He shown in Figure \ref{fig_cc_H2_models}.
\label{fig_cc_obs_vs_H2b}}
\end{figure*}

%%% Fig. 14: color-color diagrams (with PDRs) %%%
\begin{figure*}
\epsscale{2.2}
%\epsscale{1.0}
\plotone{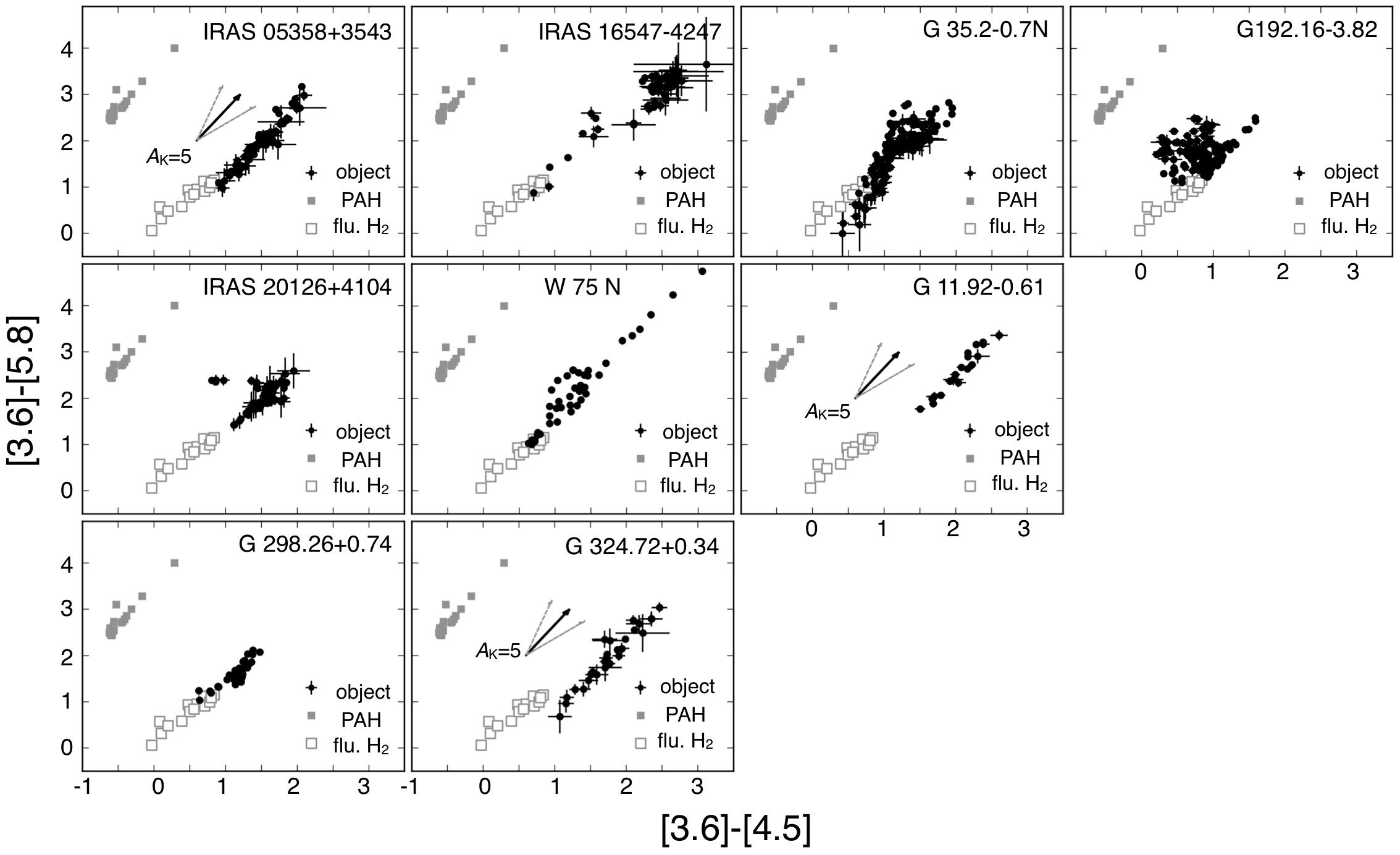}
%\plotone{f14.eps}
\caption{
Same as Figure 4 but with models for PAH and fluorescent H$_2$ emission. See text for details of models.
\label{fig_cc_with_PDRs}}
\end{figure*}

%% This figure uses \includegraphics to scale and rotate the still frame

%% If you are not including electonic art with your submission, you may
%% mark up your captions using the \figcaption command. See the
%% No more than seven \figcaption commands are allowed per page,
%% so if you have more than seven captions, insert a \clearpage

%% Tables should be submitted one per page, so put a \clearpage before

%% Two options are available to the author for producing tables:  the
%% deluxetable environment provided by the AASTeX package or the LaTeX
%% table environment.  Use of deluxetable is preferred.
%%

%% Three table samples follow, two marked up in the deluxetable environment,
%% one marked up as a LaTeX table.

%% In this first example, note that the \tabletypesize{}
%% command has been used to reduce the font size of the table.
%% We also use the \rotate command to rotate the table to
%% landscape orientation since it is very wide even at the
%% Note also that the \label command needs to be placed

%% This table also includes a table comment indicating that the full
%% version will be available in machine-readable format in the electronic

\clearpage

%%% Table 1: Line list %%%

\begin{deluxetable}{lcrc}
%\vspace{2cm}
\tabletypesize{\scriptsize}
\tablecolumns{17} 
\tablewidth{0pc} 
\tablecaption{List of Lines in the IRAC Bands \label{line list}} 
%\scriptsize
\tablehead{
\colhead{Transition} &
\colhead{$\lambda$ (\micron)\tablenotemark{a}} &
\colhead{$E_u/k$ ($\times 10^3$K )\tablenotemark{a}} &
\colhead{IRAC Band\tablenotemark{b}}

}
\startdata

\multicolumn{4}{c}{--- Bright H$_2$ lines\tablenotemark{c} ---}  \vspace{0.1cm}\\
1--0 $O$(5)     &  3.235   &  7.0 &  1\\
2--1 $O$(5)     &  3.438   &  12.6&  1\\
1--0 $O$(6)     &  3.501   &  7.6 &  1\\
%2--1 $O$(6)     &  3.723   &  &  1\\
0--0 $S$(14)   &  3.724   &  19.4 &  1\\
1--0 $O$(7)     &  3.808   &   8.4 &  1\\
0--0 $S$(13)   &  3.846   &  17.4&  1\\
0--0 $S$(12)   &  3.997   &  15.5&  2\\
0--0 $S$(11)   &  4.181   &  13.7&  2\\
0--0 $S$(10)   &  4.410   &  11.9&  2\\
1--1 $S$(11)   &  4.417   &  19.0&  2\\
0--0 $S$(9)     &  4.695   &  10.3&  2\\
1--1 $S$(9)     &  4.953   &  15.7&  2\\
0--0 $S$(8)     &  5.053   &  8.7&  3\\
0--0 $S$(7)     &  5.511   &  7.2&  3\\
%1--1 $S$(7)     &  5.810   &  &  3\\
0--0 $S$(6)     &  6.109   & 5.8 &  3\\
0--0 $S$(5)     &  6.909   & 4.6 &  4\\
1--1 $S$(5)     &  7.281   & 4.6 &  4\\
0--0 $S$(4)     &  8.026   & 3.5 &  4\\
0--0 $S$(3)     &  9.665   & 2.5 &  4\vspace{0.1cm} \\

\multicolumn{4}{c}{--- Possible atomic and ionic lines in shocks\tablenotemark{d} ---}  \vspace{0.1cm}\\
%							% upper energy level (cm-1)
Pf $\gamma$ 	& 3.741	&  155.3 & 1\\	% 107965.0568
Br $\gamma$ 	& 4.052	&  151.5 & 2\\	% 105291.65
Pf $\beta$ 	& 4.654 	&  154.6 & 2\\	% 107440.4508
Fe II 			& 5.339 	&  2.7 & 3\\	% 1872.567
Ni II 			& 6.636 	&  2.2 & 4\\	% 1506.94
Ar II 			& 6.985 	&  2.1 & 4\\	% 1431.5831
Pf $\alpha$ 	& 7.460 	&  153.4 & 4\\	% 106632.16
Ar III 			& 8.991 	& 1.6  & 4\\	% 1112.17

\enddata 

%\tableline

%---

%\end{tabular}

\tablenotetext{a}{Based on \citet{DB96} for H$_2$,and  National Institute of Standards and Technology (NIST) Atomic Spectra Data Base (http://www.nist.gov/pml/data/asd.cfm) for atomic and ionic lines.}
\tablenotetext{b}{1--4 for 3.6, 4.5, 5.8, and 8.0 \micron, respectively. See, e.g., \citet{Reach06} for the spectral response function.}
\tablenotetext{c}{The line list is based on \citet{Smith05a,Ybarra09} and calculations by \citet{Takami10b}. Note that the four IRAC filters cover more than 200 transitions of molecular hydrogen, and those not listed here significantly contribute to the flux in the 3.6-\micron~band at $T$=3000--4000 K, i.e., close to the dissociation temperature \citep[][see also Section 4.2]{Takami10b}. Note that the 1-0 S(1) line (2.12 \micron) described in the text has an upper energy of $7.0 \times 10^3$ K, i.e., the same as 1-0 O(5) in the list.}
\tablenotetext{d}{The line list is based on \citet{Reach06}. Note that the the line fluxes highly depend on the ionization state, which is a function of temperature, density, and UV field produced in high-temperature slabs \citep[e.g.,][]{Hollenbach89}. Thus, the upper level energies do not directly indicate the temperature which the individual emission lines trace.}
%\end{scriptsize}
%\end{center}
%\end{table}
\end{deluxetable} 

\clearpage

%%% Table 2: Target List %%%

\begin{table}
\begin{center}
\caption{Target List \label{target list}}
\hspace*{-1.5cm}
\begin{scriptsize}
\begin{tabular}{lcrrccccl}
\tableline\tableline
Object & Category\tablenotemark{a} & R.A. (2000)\tablenotemark{b} & Dec (2000)\tablenotemark{b}
& Distance\tablenotemark{c} & \multicolumn{3}{c}{Uncertainty of IRAC}& Region selected for color-color \\ 
&&&&(kpc)  & \multicolumn{3}{c}{flux\tablenotemark{d} (MJy str$^{-1}$)} & diagram \\
&&&&& 3.6 \micron&4.5 \micron&5.8 \micron\\ \tableline

%IRAS 05358+3543 	& HPMO 	& 5h 39m 13.0s 	& 35$^\circ$ 45' 51"		& 1.8 	& $>$ 8.0 MJy str$^{-1}$ (15-$\sigma$) for 4.5 \micron\\
%IRAS 16547--4247 	& HPMO 	& 16h 58m 17.2s	& --42$^\circ$ 52' 08"	& 2.9 	& $>$ 22 MJy str$^{-1}$ (15-$\sigma$) for 4.5 \micron\\
%G 35.2--0.7 N		& HC HII 	& 18h 58m 13.0s	& 1$^\circ$ 40' 36"		& $\sim$2 & $>$ 6.3 MJy str$^{-1}$ (15-$\sigma$) for 4.5 \micron\\
%G 192.16--3.82		& HC HII 	& 5h 58m 13.5s	& 16$^\circ$ 31' 58"	& $\sim$2 	& $>$ 2.4 MJy str$^{-1}$ (15-$\sigma$) for 4.5 \micron\\
I%RAS 20126+4104	& HC HII 	& 20h 14m 26.0s	& 41$^\circ$ 13' 33"		& 1.7 	& $>$ 14 MJy str$^{-1}$ (15-$\sigma$) for 4.5 \micron, $>$ 17 MJy str$^{-1}$ (5-$\sigma$) for 5.8 \micron\\
%                                                   										&&&&	& $>$ 17 MJy str$^{-1}$ (5-$\sigma$) for 5.8 \micron\\
%G 5.89--0.39		& UC HII 	& 18h 00m 30.4s	& --24$^\circ$ 04' 02"	& $\sim$2 & $>$ 45 MJy str$^{-1}$ (15-$\sigma$) for 4.5 \micron\\
%W 75 N			& UC HII 	& 20h 38m 36.4s	& 42$^\circ$ 37' 34"		&  $\sim$2 & $> 1.0 \times 10^2$ MJy str$^{-1}$ for 4.5 \micron\\
%G 11.92--0.61		& C08 	& 18h 13m 58.1s	& --18$^\circ$ 54' 17"	& 3.8 	& $>$19 MJy str$^{-1}$ (15-$\sigma$) for 4.5 \micron\\
%G 298.26+0.74		& C08 	& 12h 11m 47.7s	& --61$^\circ$ 46' 21"	& --- 		& $>$3.2 MJy str$^{-1}$ (80-$\sigma$) for 4.5 \micron\\
%G 324.72+0.34		& C08	& 15h 34m 57.5s	& --55$^\circ$ 27' 26"	& --- 		& $>$ 4.2 MJy str$^{-1}$ (15-$\sigma$) for 4.5 \micron\\

IRAS 05358+3543 	& HPMO 	& 5h 39m 13.0s 	& 35$^\circ$ 45' 51"		& 1.8 	&0.44&0.53&0.99& $>$15-$\sigma$ for 4.5 \micron\\
IRAS 16547--4247 	& HPMO 	& 16h 58m 17.2s	& --42$^\circ$ 52' 08"	& 2.9 	&0.67&0.98&1.44& $>$15-$\sigma$ for 4.5 \micron\\ 
G 35.2--0.7 N		& HC HII 	& 18h 58m 13.0s	& 1$^\circ$ 40' 36"		& $\sim$2 &0.36&0.42&0.87& $>$15-$\sigma$ for 4.5 \micron\\
G 192.16--3.82		& HC HII 	& 5h 58m 13.5s	& 16$^\circ$ 31' 58"	& $\sim$2 	&0.14&0.16&0.46& $>$15-$\sigma$ for 4.5 \micron\\
IRAS 20126+4104	& HC HII 	& 20h 14m 26.0s	& 41$^\circ$ 13' 33"		& 1.7 	&0.50&0.94&3.45& $>$15-$\sigma$ for 4.5 \micron, $>$5-$\sigma$ for 5.8 \micron\\
%                                                   										&&&&	& $>$ 
%G 5.89--0.39		& UC HII 	& 18h 00m 30.4s	& --24$^\circ$ 04' 02"	& $\sim$2 & $>$ 15-$\sigma$ for 4.5 \micron\\
W 75 N			& UC HII 	& 20h 38m 36.4s	& 42$^\circ$ 37' 34"		&  $\sim$2 &0.28&0.19&0.57& $> 1.0 \times 10^2$ MJy str$^{-1}$ for 4.5 \micron\\
G 11.92--0.61		& C08 	& 18h 13m 58.1s	& --18$^\circ$ 54' 17"	& 3.8 	&0.43&1.24&0.77 & $>$15-$\sigma$ for 4.5 \micron\\
G 298.26+0.74		& C08 	& 12h 11m 47.7s	& --61$^\circ$ 46' 21"	& --- 		&0.23&0.22&0.24& $>$80-$\sigma$ for 4.5 \micron\\
G 324.72+0.34		& C08	& 15h 34m 57.5s	& --55$^\circ$ 27' 26"	& --- 		&0.20&0.28&0.40& $>$15-$\sigma$ for 4.5 \micron\\

\tableline
\end{tabular}
\end{scriptsize}
\tablenotetext{a}{HPMO ... high-mass protostellar objects without any previous detection of hypercompact or
                               ultracompact H II regions; HC/UC HII ... hypercompact/ulracompact H II regions; C08 ... candidates of high-mass
                               protostars discovered in Spitzer GLIMPSE survey by Cyganowski et al. (2008). The categorization
                               of HPMO, HC H II and UC H II is based on \citet[][]{Beuther05}.}
\tablenotetext{b}{Positions for HPMO, HC/UC H II are based on millimeter interferometry by \citet{Beuther02, Franco09, Gibb03, Shepherd00, Shepherd01_science, Shepherd03, Sollins04}. Those for the three additional objects are from \citet{Cyganowski08}.  }
%\tablenotetext{c}{References --- Beuther et al. (2002, 2003); Rodr\'iguzes et al. (2008); Brand \& Blitz (1993), Gibb et al. (2003), Shepherd et al. (2000, 2001), Velaquez et al. (2002), Dickel et al. (1969). }
\tablenotetext{c}{Quoted from  \citet{Cyganowski09} for G 11.92--0.61, and the above references for the others. Note that our analysis and discussion are independent of the distance listed here, as seen in Sections 3--6. }
\tablenotetext{d}{Measured at an angular resolution of $\sim$3" (see text).}
\end{center}
\end{table}

\clearpage

%%% Table 3: fluorescent H2 and PAH %%%

\vspace{2cm}
\begin{deluxetable}{rrrrrrrrccccccccc} 
\vspace{2cm}
\rotate
\tabletypesize{\scriptsize}
\tablecolumns{17} 
\tablewidth{0pc} 
\tablecaption{IRAC Flux Estimated for fluorescent H$_2$ and PAHs\tablenotemark{a} (MJy str$^{-1}$)\label{tbl-fluH2-PAH}} 
\scriptsize
\tablehead{
\colhead{} &
\multicolumn{6}{c}{Fluorescent H$_2$} &
\colhead{} &
\multicolumn{9}{c}{PAH} \\
\cline{2-7} \cline{9-17} \\
\colhead{$\chi$\tablenotemark{b}} &
\colhead{$T_0$} &
\colhead{$n_H$} &
\colhead{} &
\colhead{} &
\colhead{} &
\colhead{} &
\colhead{} &
\multicolumn{4}{c}{$q_{PAH}$=0.47\tablenotemark{c}} &
\colhead{} &
\multicolumn{4}{c}{$q_{PAH}$=4.58\tablenotemark{c}} \\
\cline{9-12} \cline{14-17} \\
\colhead{} &
\colhead{(K)} &
\colhead{(cm$^{-3}$)} &
\colhead{3.6 \micron} &
\colhead{4.5 \micron} &
\colhead{5.8 \micron} &
\colhead{8.0 \micron} &
\colhead{} &
\colhead{3.6 \micron} &
\colhead{4.5 \micron} &
\colhead{5.8 \micron} &
\colhead{8.0 \micron} &
\colhead{} &
\colhead{3.6 \micron} &
\colhead{4.5 \micron} &
\colhead{5.8 \micron} &
\colhead{8.0 \micron}
}
\startdata 
1.0$\times 10^2$&	1000&1.0$\times 10^3$	&0.03	&0.02	&0.01	&0.10	&&0.25			&8.7$\times 10^{-2}$&0.96			&4.3				&&3.2			&1.1				&11				&38		\\
1.0$\times 10^2$&	500	&1.0$\times 10^3$	&0.02	&0.01	&$<$0.01	&0.01	&&0.25			&8.7$\times 10^{-2}$&0.96			&4.3				&&3.2			&1.1				&11				&38		\\
1.0$\times 10^2$&	300	&1.0$\times 10^4$	&0.03	&0.02	&0.01	&0.01	&&0.25			&8.7$\times 10^{-2}$&0.96			&4.3				&&3.2			&1.1				&11				&38		\\
1.0$\times 10^2$&	500	&1.0$\times 10^4$	&0.04	&0.02	&0.01	&0.04	&&0.25			&8.7$\times 10^{-2}$&0.96			&4.3				&&3.2			&1.1				&11				&38		\\
1.0$\times 10^3$&	1000	&1.0$\times 10^4$	&0.21	&0.16	&0.08	&0.70	&&2.5			&0.89			&11				&62				&&32			&11				&1.2$\times 10^2$	&4.0$\times 10^2$	\\
1.0$\times 10^3$&	500	&1.0$\times 10^4$	&0.18	&0.13	&0.05	&0.06	&&2.5			&0.89			&11				&62				&&32			&11				&1.2$\times 10^2$	&4.0$\times 10^2$	\\
1.0$\times 10^3$&	300	&1.0$\times 10^5$	&0.27	&0.21	&0.10	&0.09	&&2.5			&0.89			&11				&62				&&32			&11				&1.2$\times 10^2$	&4.0$\times 10^2$	\\
1.0$\times 10^3$&	500	&1.0$\times 10^5$	&0.26	&0.23	&0.11	&0.27	&&2.5			&0.89			&11				&62				&&32			&11				&1.2$\times 10^2$	&4.0$\times 10^2$	\\
1.0$\times 10^4$&	1000	&1.0$\times 10^5$	&0.98	&0.94	&0.48	&1.73	&&26			&9.9				&1.7$\times 10^2$	&1.3$\times 10^3$	&&3.3$\times 10^2$	&1.2$\times 10^2$	&1.3$\times 10^3$	&4.8$\times 10^3$	\\
1.0$\times 10^4$&	500	&1.0$\times 10^5$	&1.01	&0.86	&0.38	&0.37	&&26			&9.9				&1.7$\times 10^2$	&1.3$\times 10^3$	&&3.3$\times 10^2$	&1.2$\times 10^2$	&1.3$\times 10^3$	&4.8$\times 10^3$	\\
1.0$\times 10^4$&	300	&1.0$\times 10^6$	&1.22	&1.19	&0.49	&0.29	&&26			&9.9				&1.7$\times 10^2$	&1.3$\times 10^3$	&&3.3$\times 10^2$	&1.2$\times 10^2$	&1.3$\times 10^3$	&4.8$\times 10^3$	\\
1.0$\times 10^4$&	500	&1.0$\times 10^6$	&1.07	&1.12	&0.46	&0.52	&&26			&9.9				&1.7$\times 10^2$	&1.3E+09			&&3.3$\times 10^2$	&1.2$\times 10^2$	&1.3$\times 10^3$	&4.8$\times 10^3$	\\
1.0$\times 10^5$&	1000	&1.0$\times 10^6$	&2.61	&2.88	&1.32	&2.32	&&3.3$\times 10^2$	&2.6$\times 10^2$	&5.0$\times 10^3$	&3.1$\times 10^4$	&&3.8$\times 10^3$	&1.5$\times 10^3$	&1.7$\times 10^4$	&6.6$\times 10^4$	\\
1.0$\times 10^5$&	500	&1.0$\times 10^6$	&3.03	&3.20	&1.43	&0.94	&&3.3$\times 10^2$	&2.6$\times 10^2$	&5.0$\times 10^3$	&3.1$\times 10^4$	&&3.8$\times 10^3$	&1.5$\times 10^3$	&1.7$\times 10^4$	&6.6$\times 10^4$	\\

\enddata 
\tablenotetext{a}{The values tabulated here are based on Section 7 of \citet{DL07}, without throughput corrections (6 \%, 16 \%, 55 \% and 39 \% for 3.6, 4.5, 5.8, and 8.0 \micron, respectively) described in Spitzer Space Telescope Observer's Manual Version 8.0.}
\tablenotetext{b}{UV radiation field normalized by averaged interstellar field}
\tablenotetext{c}{Abundance of PAHs compared with the entire mass of carbon grains (percent) }
\end{deluxetable}

%\clearpage

%% Tables may also be prepared as separate files. See the accompanying
%% sample file table.tex for an example of an external table file.
%% To include an external file in your main document, use the \input
%% command. Uncomment the line below to include table.tex in this
%% sample file. (Note that you will need to comment out the \documentclass,
%% \begin{document}, and \end{document} commands from table.tex if you want
%% to include it in this document.)


%% The following command ends your manuscript. LaTeX will ignore any text

\end{document}